\documentstyle[12pt] {article}
\newcommand{\be}{\begin{equation}}
\newcommand{\eeq}{\end{equation}}

\def\half{{\textstyle{1\over2}}}
\def\quarter{{\textstyle{1\over4}}}

\def\p1half{{\textstyle{{{p+1}\over{2}}}}}

\def\23phalf{{\textstyle{{{23-p}\over{2}}}}}

\def\quarter{{\textstyle{1\over4}}}

\def\2p{2\pi\alpha^{\prime}}

\def\8p{8\pi^2\alpha^{\prime}}

\def\half{{\textstyle{1\over2}}}

\def\quarter{{\textstyle{1\over4}}}

\def\half{{\textstyle{1\over2}}}
\def\quarter{{\textstyle{1\over4}}}

\def\p1half{{\textstyle{{{p+1}\over{2}}}}}

\def\23phalf{{\textstyle{{{23-p}\over{2}}}}}

\def\quarter{{\textstyle{1\over4}}}

\csname @addtoreset\endcsname{equation}{section}

\def\2p{2\pi\alpha^{\prime}}

\def\8p{8\pi^2\alpha^{\prime}}

\def\d0{{\rm D0brane}}

\def\half{{\textstyle{1\over2}}}

\def\quarter{{\textstyle{1\over4}}}

\begin{document}
\thispagestyle{empty}
\begin{titlepage}
\bigskip
\hskip 3.7in{\vbox{\baselineskip12pt
}}

\centerline{\large\bf Euclidean Time Formulation for the Superstring Ensembles:}
\medskip
\centerline{\large\bf Perturbative Canonical Ensemble with Neveu-Schwarz B Field Backgrounds}
\bigskip
\bigskip
\bigskip\bigskip
\bigskip\bigskip 
\centerline{\bf Shyamoli Chaudhuri$^\dagger$\footnote{Email: scplassmann@gmail.com} }

\bigskip\bigskip
\centerline{Princeton University Libraries}
\centerline{Princeton, NJ 08544}
\bigskip

\bigskip

\date{today}
\begin{abstract}
\vskip 0.1in \noindent The lengthy discussion and derivations given in this paper have been corrected, and completed with
greater simplicity. We show that the high temperature limits of the heterotic $E_8$$\times$$E_8$ and ${\rm Spin 32}/{\rm Z_2}$ strings and their Type I A/B superstring duals are finite and convergent. The Hagedorn growth of the degeneracies in the string mass level expansion is suppressed by an exponential linear in the mass level number for both heterotic strings, and suppressed by the exponential of the negative square root of the mass level number for the Type IB superstring. However, in the massless gauge field theoretic limit of the Type IB open and closed superstring, we find clear evidence for the thermal deconfinement phase transition at the self-dual temperature by examining the annulus graph alone. Above the self-dual temperature, there is a discontinuity in the first derivative with respect to temperature of both the free energy, and the heavy quark potential, leading to a deconfined thermal gluon ensemble, with universal $1/r$ potential, and temperature dependent corrections, as predicted by Luscher and Weisz. A number of essential aspects of the worldsheet formalism of the heterotic strings are derived in an appendix, deducing thereby the O8-D0-D8brane Type IA duals of all of the heterotic CHL island universe moduli spaces. 
\end{abstract}

\end{titlepage}   

\section{Introduction}

\vskip 0.1in \noindent Nonabelian gauge theories have rich vacuum dynamics, the subject of extensive continuum and lattice investigations since the discovery of their relevance to the strong interactions \cite{willat}. Perturbative QCD is spectacularly successful as a renormalizable weakly coupled theory of the pointlike interactions of quarks and gluons \cite{pert}, and can be directly compared with calculations in the low energy gauge theory limit of the open string sector of the type IB/IA superstring theories \cite{eereview}, a subject we will not have space to examine in this paper. Lattice gauge theory (LGT) \cite{willat,peskin} in principle provides a gauge invariant formalism for short distance QCD vacuum phenomena \cite{kuti}, and has given support to the effective string analyses. 

\vskip 0.1in \noindent 
The Nambu-Goto-Eguchi-Schild string has long been the prototype effective string theory description of QCD flux lines and flux tubes \cite{effective,devecchia,luscher,alvarez} and it is well-known that the Nambu-Goto and Polyakov string actions are classically equivalent. While there has been an explosion of work applying supergravity-M theory techniques to the study on nonperturbative strongly coupled supersymmetric large N gauge theories in recent years using Maldacena's gauge-gravity dualities \cite{malda}, our current investigation examines instead the anomaly-free perturbative nonabelian gauge theories \cite{bgs} derived from exactly known one-loop results for type I/heterotic superstring theory amplitudes--- focusing in particular on the phase structure of the finite temperature nonabelian gauge theories appearing in the low energy field theory limit of the open string sector of the type IB/IA superstrings.  

\vskip 0.1in \noindent 
Our starting point is finite temperature string theory in the Euclidean time formulation, compactifying the 10D N=1 superstring theory, heterotic or Type I, on the twisted 2-torus, where Euclidean time is the circle of radius $\beta$$=$$1/T$, and the $B$-field: $B$$=$$B_{09}$$=$${\rm tanh} (\pi \alpha)$, $\alpha = (\beta_C/\beta)$, is linear in the temperature, at low temperatures, asymptoting to unity at the self dual point. The twist achieves the necessary physical gauge condition on the finite temperature quantization of the (Neveu-Schwarz) two-form gauge potential which couples to the fundamental string, in every superstring theory, heterotic, or type IA-B, or type IIA-IIB. This is the analog of the axial gauge choice on the finite temperature Yang-Mills one-form gauge potential, $A_0$$=$ constant \cite{gyp,bernard}. The 2-torus is a complex manifold, and marginal deformations are parametrized by two real numbers\footnote{As has been noted by Aspinwall for $K3$ compactifications \cite{aspinotes}, a constant, and periodic, $B$-field on a K3 surface cannot be strictly taken to zero, without approaching a singularity at finite distance in the moduli space. The non-vanishing $B$ field is necessary in order to approach several of the enhanced gauge symmetry points of interest, related to the orbifold points on the K3 moduli space. This is not to be confused with the $B$-field on the $T_9$-dualized 2-torus, $S^1$$\times$$S^1/{\rm Z}_2$, where the constant $B$-field, $B_{09}$$=$$-B_{90}$, must vanish in the supersymmetric large radius limits.} :
\begin{equation}
B = \half B_{i \bar j}  dx^i \wedge d x^{\bar j} , \quad J = \half g_{ i \bar j} dx^i \wedge d x^{\bar j} \quad ,
\end{equation}
and shifts in $B$, $J$, appear in the mass level expansions thru the, thermal and spatial, momentum modes and winding modes. Shifts of B by an integer leave the action, and all correlation functions, unchanged, due to the$SL(2,{\rm Z})$ symmetry generated by $ \sigma = {{1}\over{4\pi\alpha^{\prime}}} (B+iJ)$. The moduli space of marginal deformations is the group $SL(2,{\rm Z})$$\times$$ SL(2,{\rm Z})$, where the additional $SL(2,{\rm Z}$ describes the complex structure of the torus: we have a rectangular domain of lengths, $R_0$, $R_9$, and $\eta $$=$$ i R_9/R_0$, and $\sigma $$=$$ {{i}\over{\alpha^{\prime}}} R_9R_0$, divide the complex plane by translations, $2 \pi R_0$, $2\pi R_9$. The thermal duality transformation, $\beta$$\leftrightarrow$$1/\beta$, quite remarkably,  can be identified as an abelian subset of the mirror map \cite{aspinotes} for the 2-torus, the simplest Calabi-Yau manifold of complex dimension one. Here, $\beta $$=$$R_0/2\pi$. Thus, $R_0$$ \leftrightarrow $$1/R_0$ generates the $Z_2$ mirror map exchanging the two $SL(2,{\rm Z})$ factors. The additional $Z_2$ symmetry is generated by complex conjugation, interchanging the two real coordinates, plus a change in the sign of $B$, namely, inversion in the upper half plane, $(\sigma, \eta) $$ \leftrightarrow $$ (-\sigma , -\eta)$.

\vskip 0.1in \noindent 
Supersymmetry is spontaneously broken at low temperatures by the graviton and gaugino masses, provided by a scalar Kahler modulus, the inverse of the radius of Euclidean time. It should be noted that supersymmetry breaking occurs at very low temperatures, quite distinct from the thermal duality transition at the self dual temperature. Nevertheless, our work provides a continuous parameterization in terms of $(\beta, B, R)$, for both of these phenomena. It should be emphasized that nothing changes if the D=10 N=1 superstring theories analyzed here are replaced by, for example, an orbifold compactification, yielding a D=4 N=1 superstring vacuum state. Our analysis can thus be said to provide direct evidence by demonstration in favor of the \lq\lq continuity" conjecture made by Poppitz, Schafer, and Unsal  \cite{poppitz}, that low temperature soft supersymmetry breaking in an N=1 nonabelian gauge theory mediated by a gaugino mass--- what has been named the SYM* theory in \cite{poppitz}--- can be seen to be continuously connected to the thermal deconfinement phase transition in thermal Yang-Mills gauge theory. The results we present here would likely hold for any K3 compactification \cite{aspinotes}, for example, on $R^3$$\times$$K3$$\times$$(S^1$$\times$$S^1/Z_2)$, where the ${\rm Z}_2$ is chosen so as to give a 4D N=1 supersymmetric gauge theory with four supercharges for generic radii and $B$-field on the $(S^1$$\times$$S^1/{\rm Z}_2)$, since they hold at the orbifold points of the K3.

\vskip 0.1in \noindent 
We begin in Section 2 by establishing the finiteness of the finite temperature one loop vacuum energy density of the heterotic $E_8$$\times$$E_8$ and ${\rm Spin 32}/{\rm Z_2}$ strings in both the low temperature supergravity and Yang Mills field theoretic limit, and in the high temperature limit, both at, and beyond, the self-dual temperature, $T_C$. We show in particular how to complete the integral over both worldsheet moduli of the one-loop torus vacuum graph of closed superstrings, correcting some errors in previous papers that might mislead the reader, a tour-de-force that has important implications for the applications of string scattering amplitudes to problems in particle and astro-particle physics \cite{eereview}. An analogous demonstration is carried out in Section 3 for the open unoriented and closed string  vacuum graphs of the Type IB superstring theory, except that the suppression of Hagedorn growth is by an exponential of the square root of the mass level number at the self-dual temperature. For clarity, we show that the thermal duality transition of the full string theory is benign: in both the heterotic and type I string theories, it is a Kosterlitz-Thouless phase transition characterized by an infinite tower of finite, and analytic, thermodynamic potentials, including the Helmholtz free energy, Gibbs free energy, entropy and specific heat. 

\vskip 0.1in \noindent 
In \cite{pairb,pairf,ncom}, we formulated the Polyakov string path integral prescription \cite{polyakov,schwarz,poltorus,cmnp} for macroscopic incoming and outgoing string states at finite separation in an embedding target spacetime. ${\cal M}_{\rm IB}  ({\bf x} )$ is the expectation value in the type IB string theory for the insertion of a macroscopic boundary loop on the worldsheet mapped to a fixed loop ${\cal C}$ at location ${\bf x}$ in the embedding target spacetime. The mapping must preserve the worldsheet super Diffeomorphism$\times$Weyl invariances, both in the bulk, and on the boundaries, of the worldsheet, due to its fixed spatial location. The observable ${\cal M}_f ({\cal C}) $ transforms in the fundamental representation $f$ of the nonabelian gauge group, and the trace over the Chan-Paton index ensures that the observable is gauge invariant. Since the endpoints of the open strings carry color charge, the boundary of the hole in the worldsheet is mapped to a Wilson loop in the low energy nonabelian gauge theory limit, describing the world-history of an infinitely massive probe carrying color charge. In Section 4, we analyze the insertion of a pair of spacelike macroscopic loop observables, mapped to the fixed spacelike loops in target spacetime, ${\cal C}_2$, ${\cal C}_1$, spatially separated by distance $R$, directly yields the potential between two massive charged color sources in the gauge field theory limit--- rather than the exponentiated potential which appears in the corresponding Wilson loop two-point function of the nonabelian gauge theory, as in \cite{poppitz}. For open string end-point Chan-Paton wave functions transforming in the fundamental representation ${ f}$ of the gauge group, we shall derive an expression for the macroscopic pair correlation function. In this analysis, we are suppressing the full content of the unoriented open and closed type I superstring theories, in order to draw attention to the properties of the massless gauge theory limit in and of itself.\cite{pairf}\footnote{This remarkable feature of Type IA-IB open and closed superstrings follows from the relation of the couplings; at tree level, it is simply $g_{\rm closed} $ $=$ $g_{\rm open}^2$. We see that the annulus with macroscopic loops can be analyzed in the nonabelian gauge theory alone, as with the finite temperature vacuum energy density, since the supergravity is decoupled at tree level. The supergravity multiplet belongs in the closed string sector of the Type IB-IA superstrings. Upon computing the renormalized couplings and string mass scale, this tree relation receives loop corrections  \cite{ncom,eereview}. Very recent progress has been made in the computation of unambiguous multi-loop superstring amplitudes, two-loop and beyond, which lends further promise to the systematic extension of our worldsheet analysis.}:
\begin{equation}
 {\cal W}^{(2)}_{\rm IB} (R) = \left < {\rm Tr}_{ f} {\cal M}_{ f}  ({\cal C}_2 ) {\cal M}_{ f} ({\cal C}_1 ) \right >  = - \beta V(R,\beta) \quad ,
\end{equation}
which interpolates neatly between the low temperature, large spatial separation, confinement regime of the nonabelian gauge theory, derived from the $T_9$, and $T_0$, dual, type IA string, and dominated by a linear term in the heavy quark potential, and the high temperature, small spatial separation, deconfinement regime, derived from the type IB string, which is dominated by the $1/R$ Luscher potential. We will show that the expression for the heavy quark potential we derive is in qualitative agreement with both the original Cornell phenomenological potential model \cite{cornell}, with the Nambu-Goto-Eguchi-Schild-Polyakov QCD effective strings \cite{alvarez,luscher,devecchia,newlw}, and with lattice gauge theory measurements \cite{peskin,kuti}. At low type IB temperatures, and for large type IB spatial separations of the infinitely heavy quarks, in addition, we derive from the expectation value of two Polyakov-Susskind loops the next-to-leading order thermal (field-dependent) corrections to the universal $1/R$ potential in the deconfined phase, thereby confirming Luscher and Weise's conjecture that the leading correction to the universal $1/R$ term is $O(1/R^3)$ \cite{luscher,newlw}. 

\vskip 0.1in \noindent 
The appendix contains a series of significant developments in the world sheet formalism of the heterotic string theories that lend insight to the results in this paper, and on the string/M theory strong-weak duality web more generally. In particular, we deduce in Appendix A.4 the Type IA strong coupling duals of the heterotic CHL orbifold island universes, using the formalism for the gauge group in O8-D0-D8brane compactifications, with 16 pairs of D0-D8branes and their images at each of two orientifold planes at the endpoints of the interval in $(S^1$$\times$$S^1/{\rm Z}_2)$ compactifications of the Type IA $O(16)$$\times$$(16)$ superstring \cite{flux}. The spinor of $O(16)$ is given by the solitonic fundamental strings created at the intersection of D0-branes with D8branes \cite{bachas,flux}, and the configuration we suggest gives the full gauge group $E_8$$\times$$E_8$. It is straightforward to then identify all of the type IA duals of the CHL orbifolds of the $E_8$$\times$$E_8$ heterotic string, and breaking the supersymmetry further by orbifold compactification gives three generations of chiral fermions in the Standard Model embedded within the spinor of $SO(16)$, a technique well-known to string and grand unified theory phenomenologists \cite{cchl,chl}.

\section{Finite temperature Heterotic String Vacuum Functional}

\vskip 0.1in \noindent The finite temperature one-loop vacuum functional 
of the $\rm E_8 \times E_8$ heterotic string is given in the Euclidean time prescription by the 
compactification of the heterotic string on the twisted 2-torus with radii, $(\beta_H, R_H)$, and constant background $B$ field parametrized as, 
$B_{09} = -B_{90} = |B|$$=$${\rm tanh } (\pi \alpha)$, where $\alpha$$=$$T/T_C$, and $T_C$ is the self-dual temperature:
\begin{eqnarray}
W_{\rm H} (\beta) =&& {\cal N} \beta_H (2 \pi R_H) L^{8} (4\pi^2 \alpha^{\prime})^{-5} \int_{\cal F} {{d^2 \tau}\over{4\tau_2^2}} \cdot 
  (\tau_2)^{-4} [\eta(\tau) {\bar{\eta}} ({\bar{\tau}} )]^{-6}      \left [ {{ e^{\pi \tau_2 \alpha^2 } \eta (\tau) }\over{ \Theta_{11} (\alpha , \tau ) }} \right ]
  \left [ {{ e^{\pi \tau_2 \alpha^2 } \eta ({\bar \tau} ) }\over{ \bar{\Theta}_{11} (\alpha , {\bar \tau} ) }} \right ] 
  \cr
  &&\quad\quad \times 
  {{1}\over{4}}  \left [ {{ {\bar{\Theta}}_{00} (\alpha, \bar\tau) }\over{ e^{\pi \tau_2 \alpha^2} {\bar{\eta}} }} \left ({{{\bar{\Theta}}_{00}}\over{{\bar{\eta}}}} \right )^3  
  - {{ {\bar{\Theta}}_{01} (\alpha, \bar\tau) }\over{ e^{\pi \tau_2 \alpha^2} {\bar{\eta}} }}\left ({{{\bar{ \Theta}}_{01}}\over{{\bar{\eta}}}} \right )^3 
  - {{ {\bar{\Theta}}_{10} (\alpha, \bar\tau) }\over{ e^{\pi \tau_2 \alpha^2} {\bar{\eta}} }}\left ({{ {\bar{\Theta}}_{10}}\over{{\bar{\eta}}}} \right )^3  \right ]   \cr
&& \quad\quad\quad \quad \quad \times  {{1}\over{4}} \left [  \left ({ { \Theta_{00} }\over{ \eta }} \right )^{8}
 + \left ({ { \Theta_{01} }\over{\eta }} \right )^{8} + 
\left ({ { \Theta_{10} }\over{ \eta }} \right )^{8}
 \right ]^2 \cr
 && \quad\quad \times \sum_{n_0,w_0=-\infty}^{\infty} \sum_{n_9,w_9=-\infty}^{\infty} \exp \left [ - \pi \tau_2 \left ( {{4\pi^2 \alpha^{\prime} n_0^2 }\over{\beta_H^2}}
             +  {{\alpha^{\prime} n_9^2 }\over{R_H^2}} \right ) \right ]  \cr              
      &&         
          \quad \quad \quad \times  \exp \left [ - \pi \tau_2 ( 1 + {\rm tanh} (\pi \alpha))^2 \left ( {{w_9^2 \beta_H^2 }\over{4\pi^2 \alpha^{\prime} }}  +
                                      {{w_0^2 R_H^2 }\over{ \alpha^{\prime} }} \right )  \right ] \cr
&&\quad \quad  \quad \quad \quad 
            \times \exp \left [ i \pi \tau_1 (n_0w_9+n_9 w_0) (1+{\rm tanh} (\pi \alpha) ) \right ] \quad .  \cr
&& \label{eq:finlat}
\end{eqnarray}
Note the presence of the holomorphic one loop $\rm E_8 \times E_8$ vacuum functional in the third line of this formula, which we leave unchanged by a 
possible Wilson line since we wish to keep the gauge group fixed. 

\vskip 0.1in \noindent 
In the high mass level number regime as we approach the self-dual temperature, the parameters $\alpha$$\to$, $\rm tanh (\pi \alpha)$, will asymptote to unity,
giving pure numerical factors; we leave them as parameterizing the background B field in this preliminary expression valid for all temperatures and mass levels. 
Expressing the theta functions, and eta functions, in terms of mass level number, namely, positive integer powers of $|q {\bar q}|$, we can expand in this variable to obtain:
\begin{eqnarray}
W_{\rm H} (\beta) =&& {\cal N} \beta_H (2 \pi R_H) L^{8} (4\pi^2 \alpha^{\prime})^{-5} \int_{\cal F} {{d^2 \tau}\over{4\tau_2}} \cdot 
  (\tau_2)^{-5}  \sum_{m=0}^{\infty} e^{-4m\pi \tau_2 } f_{\rm E_8 \times E_8}^{(m)} (1+ {\rm tanh} (\pi \alpha) )   \cr
  && \quad \quad \times \sum_{n_0,w_0=-\infty}^{\infty} \sum_{n_9,w_9=-\infty}^{\infty} 
  \exp \left [ - \pi \tau_2 \left ( {{4\pi^2 \alpha^{\prime} n_0^2 }\over{\beta_H^2}}
             +  {{\alpha^{\prime} n_9^2 }\over{R_H^2}} \right ) \right ]  \cr              
      &&         
          \quad \quad \quad \times  \exp \left [ - \pi \tau_2 ( 1 + {\rm tanh} (\pi \alpha))^2 \left ( {{w_9^2 \beta_H^2 }\over{4\pi^2 \alpha^{\prime} }}  +
                                      {{w_0^2 R_H^2 }\over{ \alpha^{\prime} }} \right )  \right ] \cr
&&\quad \quad  \quad \quad \quad 
            \times \exp \left [ i \pi \tau_1 (n_0w_9+n_9 w_0) (1+{\rm tanh} (\pi \alpha) ) \right ] \quad .  
        \label{eq:finlatth}
\end{eqnarray}
where the $ f^{(m)} (0)$, and $f^{(m)} (2)$, are, respectively, the numerical mass degeneracies of the partition function of the 
$\rm E_8\times E_8$ heterotic string in the supersymmetric zero temperature limit, and in the vicinity of the high temperature 
self dual critical point. Solving for the Helmholtz free energy at one loop order in string perturbation theory, 
we obtain:
\begin{eqnarray}
F_{\rm H} (\beta) =&&- {\cal N} V \quarter (4\pi^2 \alpha^{\prime})^{-5} \int_{-1/2}^{1/2} d \tau_1 \int_{(1-\tau_1^2)^{1/2}}^{\infty} d \tau_2 \cr
&&\quad \quad \times   
  (\tau_2)^{-6} \sum_{m=0}^{\infty} e^{-4m\pi \tau_2 } f_{\rm E_8 \times E_8}^{(m)} (1+{\rm tanh} (\pi \alpha ) )   \cr
  && \times \sum_{n_0,w_0=-\infty}^{\infty} \sum_{n_9,w_9=-\infty}^{\infty} 
       \exp \left [ -\pi \tau_2  \left (   {{4\pi^2 \alpha^{\prime} n_0^2 }\over{\beta_H^2}}   
            +  {{\alpha^{\prime} n_9^2 }\over{R_H^2}}  \right ) \right ] \cr           
         &&   \quad \quad \quad \times  \exp \left [ - \pi \tau_2 ( 1 + {\rm tanh} (\pi \alpha))^2 \left ( {{w_9^2 \beta_H^2 }\over{4\pi^2 \alpha^{\prime} }}  +
                                      {{w_0^2 R_H^2 }\over{ \alpha^{\prime} }} \right )  \right ] \cr
             && \quad \quad \quad \times  \exp \left [ i \pi \tau_1 (n_0w_9+n_9 w_0) ( 1 + {\rm tanh} (\pi \alpha)) \right ] \quad  . 
        \label{eq:finlatthf}
\end{eqnarray}
More familiar to a particle physicist, the Helmholtz free energy at one loop order is nothing but the one-loop vacuum energy density of the finite
temperature string vacuum, $F$$=$$V_9 \rho_H $$=$$-W_H /\beta_H$. Substituting $y$$=$$1/\tau_2$, we can express the one-loop vacuum 
energy density at finite temperature as:
\begin{eqnarray}
\rho_{\rm H} (\beta) =&&- {\cal N} (4\pi^2 \alpha^{\prime})^{-5}  \sum_{m=0}^{\infty} \sum_{n_0,w_0=-\infty}^{\infty} \sum_{n_9,w_9=-\infty}^{\infty} 
f_{\rm E_8 \times E_8}^{(m)} 
            (1+{\rm tanh} (\pi \alpha ) ) \cr
            && \quad\quad \times             
             \int_{-1/2}^{1/2} d \tau_1   \exp \left [ i \pi \tau_1 (n_0w_9+n_9 w_0)  \right ] 
\int_0^{(1-\tau_1^2)^{-1/2}} d y y^{4} e^{ - A / y}    \quad ,
        \label{eq:freeh}
\end{eqnarray}
where the function in the exponent, $A$, is the mass formula for the finite temperature heterotic string spectrum:
\begin{equation}
A_{\rm H} (\beta, \alpha; m) = m + 
   \quarter \left [   {{4\pi^2 \alpha^{\prime} n_0^2 }\over{\beta_H^2}}   
            +  {{\alpha^{\prime} n_9^2 }\over{R_H^2}}  + ( 1 + {\rm tanh} (\pi \alpha))^2 \left ( {{w_9^2 \beta_H^2 }\over{4\pi^2 \alpha^{\prime} }}  +
                                      {{w_0^2 R_H^2 }\over{ \alpha^{\prime} }} \right )  \right ]   \quad . 
 \label{eq:mass}
 \end{equation}
Note that mass level number, $m$, by mass level number, there is an infinite tower of thermal momentum and thermal winding modes in the finite temperature
spectrum, in addition to the tower of possible spatial momenta and windings, as a consequence of the generalized axial gauge condition necessitating 
compactification on the Neveu-Schwarz $B_{09}$-field twisted 2-torus. 

\vskip 0.1in \noindent 
The integral over the variable $y$ in the expression for the vacuum energy density can be recognized as a standard integral representation of the 
Whittaker function, ${\cal W}_{-{{\nu + 1 }\over{2}} , {{\nu}\over{2}}}(Au)$:
\begin{eqnarray}
\rho_{\rm H} (\beta) =&& - {\cal N} (4\pi^2 \alpha^{\prime})^{-5} \sum_{m=0}^{\infty} 
 \sum_{n_0,w_0=-\infty}^{\infty} \sum_{n_9,w_9=-\infty}^{\infty} 
f_{\rm E_8 \times E_8}^{(m)} (1+{\rm tanh} (\pi \alpha ) )   \cr
&& \quad\quad \quad \times  \int_{-1/2}^{1/2} d \tau_1   \exp \left [i \pi \tau_1 (n_0w_9+n_9 w_0)  \right ]  
    A^{{\nu -1}\over{2}}  u^{{{\nu +1}\over{2}}} e^{- \half A /u} {\cal W}_{ -{{\nu + 1 }\over{2}} , {{\nu}\over{2}} } (A/ u ) 
\quad ,
\cr
&& \quad \quad \quad \quad
   {\rm where} \quad \nu = 5  ,  \quad  {\rm and }  \quad u \equiv (1-\tau_1^2)^{-1/2}  \quad ,
\label{eq:whitt}
\end{eqnarray}
and upon substitution, it is helpful to make the change of variable $x^2$$=$$1-\tau_1^2$$=$$1/u^2$. The one-loop vacuum energy density therefore takes
the form: 
\begin{eqnarray}
\rho_{\rm H} (\beta) =&& - 2 {\cal N}  (4\pi^2 \alpha^{\prime})^{-5} 
\sum_{m=0}^{\infty}  \sum_{n_0,w_0=-\infty}^{\infty} \sum_{n_9,w_9=-\infty}^{\infty} 
f_{\rm E_8 \times E_8}^{(m)} (1+{\rm tanh} (\pi \alpha ) )   \cr
&& \quad \quad  \times 
\int_{0}^{{\sqrt 3/2}} d x  ~  {\rm Cos} \left [\pi {\sqrt{1-x^2}} (n_0w_9+n_9 w_0)  \right ]  
    A^{{\nu -1}\over{2}}  x^{-{{\nu +1}\over{2}} + 1 } e^{- \half A x} {\cal W}_{ -{{\nu + 1 }\over{2}} , {{\nu}\over{2}} } (A x )  . \cr
    &&
\label{eq:whittx}
\end{eqnarray}
Alternatively, we can use the integral representation in terms of the inverse variable, $u$:
\begin{eqnarray}
\rho_{\rm H} (\beta) =&& - 2 {\cal N}  (4\pi^2 \alpha^{\prime})^{-5} 
\sum_{m=0}^{\infty}  \sum_{n_0,w_0=-\infty}^{\infty} \sum_{n_9,w_9=-\infty}^{\infty} 
f_{\rm E_8 \times E_8}^{(m)} (1+{\rm tanh} (\pi \alpha ) )  \int_{0}^{{2/\sqrt 3}} d u  ~  (1-u^{-2})^{-1/2} \cr
&& \quad\quad   \times 
 ~ {\rm Cos} \left [\pi (1-u^{-2} )^{1/2} (n_0w_9+n_9 w_0)  \right ]  
    A^{{\nu -1}\over{2}}  u^{-3+{{\nu +1}\over{2}}  } e^{- \half A /u} {\cal W}_{ -{{\nu + 1 }\over{2}} , {{\nu}\over{2}} } (A /u )  . \cr
    &&
\label{eq:whittxu}
\end{eqnarray}

\vskip 0.1in \noindent
We will be interested in the low temperature, power law, and high temperature asymptotics of the Whittaker function. Prior to that step, notice that 
the cosine function, and its argument, can be further simplified, replacing each by their Taylor expansions, both of which are completely valid 
in the domain of the integral over $x$. This is the form of the expression for the one-loop vacuum energy density at finite temperature which 
we will analyze in the low temperature limit in what follows. With this substitution, we get the expression:
\begin{eqnarray}
\rho_{\rm H} (\beta) =&& - 2 {\cal N}  (4\pi^2 \alpha^{\prime})^{-5} 
\sum_{m=0}^{\infty}  \sum_{n_0,w_0=-\infty}^{\infty} \sum_{n_9,w_9=-\infty}^{\infty} 
f_{\rm E_8 \times E_8}^{(m)} (1+{\rm tanh} (\pi \alpha ) )   \cr
&& \quad \quad  \times 
\int_{0}^{{\sqrt 3/2}} d x  ~ \left [ \sum_{s=0}^{\infty}  \sum_{l=0}^s  {{ s!}\over{ l! (s-l)!}} {{(-1)^{l+2s}}\over{(2s)!}} ~ (n_0w_9+n_9 w_0)^{2s}  (x)^{2l}    \right ]  \cr
&& \quad \quad \quad \quad \quad \times A^{2}  ~ x^{-2 } ~ e^{- \half A x} ~ {\cal W}_{ - 3 , {{5}\over{2}} } (A x )  \quad .
\label{eq:whittaylori}
\end{eqnarray}
Alternatively, we use the inverse variable, $u$. This is the form of the expression for the one-loop vacuum energy density at finite temperature which
we will analyze in the high temperature limit in what follows. With the two substitutions, we get the alternative expression:
\begin{eqnarray}
\rho_{\rm H} (\beta) =&& - 2 {\cal N}  (4\pi^2 \alpha^{\prime})^{-5} 
\sum_{m=0}^{\infty}  \sum_{n_0,w_0=-\infty}^{\infty} \sum_{n_9,w_9=-\infty}^{\infty} 
f_{\rm E_8 \times E_8}^{(m)} (1+{\rm tanh} (\pi \alpha ) )   \cr
&& \quad \quad  \times 
\int_{0}^{{2/\sqrt 3}} d u  ~  \left [ \sum_{s=0}^{\infty}  \sum_{l=0}^s  {{ s!}\over{ l! (s-l)!}} {{(-1)^{l+2s}}\over{(2s)!}} ~ (n_0w_9+n_9 w_0)^{2s}  (u)^{-2l}    \right ]  \cr
  && \quad \quad  \times 
    \sum_{r=0}^{\infty} (-1)^r {{u^{-2r}}\over{r!}} \left ( (\half) \cdot (\half \cdot 3) \cdots (r-\half)\right ) 
    A^{{\nu -1}\over{2}}  u^{-3+{{\nu +1}\over{2}}  } e^{- \half A /u} {\cal W}_{ -{{\nu + 1 }\over{2}} , {{\nu}\over{2}} } (A /u )  . \cr
    &&
\label{eq:whittxui}
\end{eqnarray}

\subsection{Low Temperature Supergravity-super-Yang Mills theory Limit}

\vskip 0.3in
\noindent We now take the low temperature field theoretic limit of this expression, verifying that it has the expected properties of a ten-dimensional finite 
temperature field theory. We substitute the power series expansion of the Whittaker function, prior to performing the integral over variable 
$x$, the $\tau_1$ worldsheet modulus. The expansion in powers of $A$, has as its leading term at low temperatures, the thermal spectrum of the massless 
modes of the supersymmetric string:
\begin{eqnarray}
\rho_{\rm H} (\beta) =&& - 2 {\cal N}  ~  {{(-1)^5 }\over{5!}} ~ (4\pi^2 \alpha^{\prime})^{-5} 
\sum_{m=0}^{\infty}  \sum_{n_0,w_0=-\infty}^{\infty} \sum_{n_9,w_9=-\infty}^{\infty} 
A^5 f_{\rm E_8 \times E_8}^{(m)} (1+{\rm tanh} (\pi \alpha ) )   \cr
&& \quad \quad\quad \quad   \times 
 ~ \left [ \sum_{s=0}^{\infty}  \sum_{l=0}^s  {{ s!}\over{ l! (s-l)!}} {{(-1)^{l+2s}}\over{(2s)!}} ~ (n_0w_9+n_9 w_0)^{2s}     \right ]  \cr 
 && \quad 
         \times ~ \{ ~ \sum_{k=0}^{\infty} A^{2l+2k}  ~  {\bf \gamma} \left ( 2 l  + k ,  \half {\sqrt 3} A  \right )  \left [ {{ \Psi (k+1) - {\rm ln} A }\over{k!}} \right ]    \cr
         && \quad \quad 
              + \sum_{k=0}^4 (-1)^{5+k} ~ \Gamma (5-k) A^{ 2l + 2k -10} ~{\bf \gamma} \left (  2 l  +k  -5 , \half {\sqrt 3}A \right ) 
          \cr
&& \quad\quad \quad    - \quad \sum_{p=1}^{\infty} \sum_{k=0}^{\infty} {{A^{2l+2k+p } }\over{ k!}} \sum_{r=0}^{p} {{ p!}\over{ r! (p-r)!}} ~ 
{\bf\gamma}  \left ( 2 l + k + p  , \half {\sqrt 3} A  \right )  ~ \}  ,
\cr
&& \label{eq:whittgama}
\end{eqnarray}
where the function $A$ gives the full thermal, and spatial, momenta and windings, of the thermal spectrum, for any given mass
level number, $m$. Expanding about the massless modes: 
\begin{equation}
A_{\rm H} ( T , \alpha ; 0 ) = 
   \quarter \left [  4\pi^2 \alpha^{\prime} n_0^2 T^2     
            +  {{\alpha^{\prime} n_9^2 }\over{R_9^2}}  + ( 1 + 2 {\rm tanh} ~ \pi (T/T_C) ) \left ( {{w_9^2  }\over{4\pi^2 \alpha^{\prime} T^2 }}  +
                                      {{w_0^2 R_9^2 }\over{ \alpha^{\prime} }} \right )  \right ]   \quad ,
 \label{eq:massh}
 \end{equation}
it is apparent that at low temperatures, the leading term is the tower of thermal Kaluza-Klein modes, or Matsubara frequencies, in the language of thermal field theories.
For low temperature, or large radius, the spatial coordinate is in the large radius limit, the spatial Kaluza Klein modes tend towards a continuum, and no spatial 
windings are excited. Thus, we first extract the thermal momentum modes, and the power $A^5$ instantly gives the expected field theoretic $T^{10}$ in the one 
loop string free energy. The correction from the tower of thermal winding and spatial Kaluza-Klein modes is a purely string theoretic artifact:
\begin{equation}
A^5 ( T , \alpha ; 0 ) ~  \simeq  ~ 
   \quarter \left (  4\pi^2 \alpha^{\prime} n_0^2 \right )^5 T^{10}  
            \times  \left \{   1 
            + 5 ~ \left \{ 1  +  {{\alpha^{\prime} n_9^2 }\over{R_9^2 4\pi^2 \alpha^{\prime } n_0^2 T^2}} + [1+ 2 \pi (T/T_C) ] {{w_0^2 R_9^2}\over{ 4\pi^2 \alpha^{\prime 2} n_0^2 T^2 }} \right \}  \right \}   \quad .
 \label{eq:masszero}
 \end{equation}
Suppressing the winding modes at the lowest temperatures, the string finite temperature vacuum energy density takes the simple form:
\begin{eqnarray}
\rho_{\rm H} (\beta) \simeq  && - 2 {\cal N}  ~ {{(-1)^5 }\over{5!}} ~ (4\pi^2 \alpha^{\prime})^{-5} 
 \sum_{n_0,w_0=-\infty}^{\infty} 
f_{\rm E_8 \times E_8}^{(0)}   \cr && \quad \quad \times 
\quarter \left (  4\pi^2 \alpha^{\prime} n_0^2 \right )^5 T^{10}  
            \times   \left \{   1 
            + 5 ~ \left [ 1 +  {{\alpha^{\prime} n_9^2 }\over{R_9^2 4\pi^2 \alpha^{\prime } n_0^2 T^2}}   \right ]  \right \}    \quad .
\label{eq:masszeroh}
 \end{eqnarray}

\subsection{High Mass Level Asymptotics at Self-Dual Temperature}

\vskip 0.3in
\noindent Finally, we substitute the asymptotic expansion of the Whittaker function in the expression derived above for the one loop free energy of the 
canonical ensemble of thermal $\rm E_8\times E_8$ heterotic strings, valid for large mass level number. The asymptotic series expands in 
negative powers of $A$, namely, large $m$, and arbitrary temperature, although we will be interested in the behavior of this sum over mass
levels in the vicinity of the self dual temperature. Note that the expression is finite, and convergent, for the full temperature range, even beyond the 
self dual temperature. We begin with the inverse variable representation:
\begin{eqnarray}
\rho_{\rm H} (\beta) =&& - 2 {\cal N}  (4\pi^2 \alpha^{\prime})^{-5} 
\sum_{m=0}^{\infty}  \sum_{n_0,w_0=-\infty}^{\infty} \sum_{n_9,w_9=-\infty}^{\infty} 
f_{\rm E_8 \times E_8}^{(m)} (1+{\rm tanh} (\pi \alpha ) )   \cr
&& \quad \quad  \times 
\int_{0}^{{2/\sqrt 3}} d u  ~  \left [ \sum_{s=0}^{\infty}  \sum_{l=0}^s  {{ s!}\over{ l! (s-l)!}} {{(-1)^{l+2s}}\over{(2s)!}} ~ (n_0w_9+n_9 w_0)^{2s}  (u)^{-2l}    \right ]  \cr
  && \quad \quad  \times 
    \sum_{r=0}^{\infty} (-1)^r {{u^{-2r}}\over{r!}} \left ( (\half) \cdot (\half \cdot 3) \cdots (r-\half)\right ) 
    A^{{\nu -1}\over{2}}  u^{-3+{{\nu +1}\over{2}}  } e^{- \half A /u} {\cal W}_{ -{{\nu + 1 }\over{2}} , {{\nu}\over{2}} } (A /u )  , \cr
    &&
\label{eq:whittxuis}
\end{eqnarray}
and substitute the asymptotic expansion for the Whittaker function, and keeping its leading term, $k$$=$$0$:
\begin{eqnarray}
F_{\rm H} (\beta) \simeq&& - 2 {\cal N} V   ~ (4\pi^2 \alpha^{\prime})^{-5} 
\sum_{m=0}^{\infty}  \sum_{n_0,w_0=-\infty}^{\infty} \sum_{n_9,w_9=-\infty}^{\infty} 
f_{\rm E_8 \times E_8}^{(m)} (2)   \cr
&& \quad \quad  \times 
 ~ \left [ \sum_{s=0}^{\infty}  \sum_{l=0}^s  {{ s!}\over{ l! (s-l)!}} {{(-1)^{l+2s}}\over{(2s)!}} ~ (n_0w_9+n_9 w_0)^{2s}     \right ] \cr 
&& \quad \quad \quad \quad  \sum_{r=0}^{\infty} (-1)^r {{1}\over{r!}} \left ( (\half) \cdot (\half \cdot 3) \cdots (r-\half)\right )  \cr 
&& \quad \quad \quad \quad \quad \quad \times \left [ 1+ \sum_{k=1}^{\infty} O(k)  \right ] \cr 
&& \times ~ \{ ~ e^{- 2A/{\sqrt 3}}  ~ A^{l+r}  (\half {\sqrt 3})^{-l-r-5/2}  {\cal W}_{4-2l-2r, 2-l-r} (2/{\sqrt 3} ) 
 \}  \quad ,  \cr
&& \label{eq:gammaf}
\end{eqnarray}
 Note that the free energy is exponentially damped as a linear power of $m$, the mass level number, correcting the Hagedorn 
growth of the numerical degeneracies as a square root of the mass level number in the vicinity of the self-dual temperature. The free 
energy is finite, and the expression given above is strongly convergent at the critical point. The function in the exponential, $A$, gives the 
full thermal, and spatial, momenta and windings, of the thermal spectrum, expanding about the asymptotic mass level number, $m$,
and setting ${\rm tanh}(\pi \alpha)$ to unity:
\begin{equation}
A_{\rm H} (\beta, \alpha; m) = m + 
   \left [  \left (  {{\pi^2 \alpha^{\prime} n_0^2 }\over{\beta_H^2}}  +  {{\alpha^{\prime} n_9^2 }\over{4 R_H^2}} 
   \right ) +   \left ( {{ w_0^2 R_H^2 }\over{ \alpha^{\prime} }}  + {{w_9^2 \beta_H^2 }\over{4 \pi^2 \alpha^{\prime} }}  \right )  \right ]   \quad , 
 \label{eq:masshg}
 \end{equation}
 where we have rearranged the formula to highlight the symmetry linking Kaluza-Klein and winding modes, the twist having mixed spatial
 and thermal coordinates.  For large $m$, and at high temperatures of order the self dual temperature, we find that all of the thermal and 
 spatial winding modes are excited, and winding modes dominate the expression for the free energy, due to the presence of the exponential,
 in the small radius high temperature limit. The $f^{m}(2)$ in the one-loop free energy are the numerical degeneracies at the critical 
 temperature $T_C$. 
 
 \vskip 0.1in \noindent
Finally, in passing, we recall that our starting point was the generating functional for connected vacuum diagrams in one loop string perturbation theory, $W$, and
we do not face the usual problems associated with taking the thermodynamic limit of the canonical partition function, $Z$. The Helmholtz free energy, $F$, 
and the finite temperature vacuum energy density, $\rho$, are related to this as follows:
\begin{equation}
W \equiv {\rm ln} Z = - \beta_H V \rho , \quad F \equiv -T_H {\rm } Z = -W/\beta_H =  V \rho , \quad P = - \left ( {{\partial F}\over{\partial V}} \right )_{T_H} = - \rho \quad ,
\label{therm}
\end{equation}
where $P$ is the pressure of the string canonical ensemble at fixed temperature, and $V$ is its spatial volume. Note that $P$ equals the negative of $\rho$ for the string canonical 
ensemble. The next few entries in the list of thermodynamic potentials are the internal or Gibbs free energy, the entropy, and the specific heat at constant volume:
\begin{equation}
U = -T_H^2 \left ( {{\partial W}\over{\partial T_H}} \right )_V , \quad S =  - \left ( {{\partial F}\over{\partial T_H}} \right )_V  , \quad C_V =  T_H \left ( {{\partial S}\over{\partial T_H}} \right )_V 
\quad . 
\label{thermp}
\end{equation}
It is evident by inspection of the expressions for the Helmholtz free energy and the detailed dependence on $\beta$ that the results for an infinity of partial derivatives are completely analytic and finite, identifying the thermal duality transition as of the Kosterlitz-Thouless type. There is no divergence in the expressions at any order in the thermodynamic potentials. We leave further discussion of this intriguing observation to the future.
 
\section{Finite temperature Type IA String Vacuum Functional}

\vskip 0.1in \noindent
We will give the analysis of the ultraviolet limit of the unoriented graphs of the O(32) type IA superstring at finite temperature. If the N=32 D8branes are all on a single O8-plane, the Dirichlet string measures the potential energy of the string stretched between the D8brane stack on an O8plane, with the O8plane defect at the other end of the interval of length, $R$. We will show at the conclusion of our derivation that it is in fact possible to take $R$ to zero, and recover the result for the finite temperature type IB O(32) vacuum, {\em without} the Dirichlet stretched string. Note that we have the constant mode of the NS sector antisymmetric tensor gauge potential, $B$, which remains after the orientation transformation which eliminates the propagation of the NS twoform field. We shall set the constant background field, $|B_{09}| = - |B_{90}| = |{\rm tanh } (\pi\alpha)| $, where $\alpha \equiv \left ( \beta_C/\beta \right ) $$=$$ \alpha^{\prime 1/2} T $, is linear in the temperature, measured in units of the inverse string scale.

\vskip 0.1in \noindent
We analyse the small $t$-- high temperature-- behavior of the three individual open and unoriented one-loop type IB superstring graphs. We begin with the result for the oriented open string sector, or annulus graph, in terms of Jacobi theta functions, where $N$ denotes the number of D8branes or, equivalently, the Chan-Paton factor carried by the endpoints of the open string. We have expressed the Jacobi theta functions in the integrand as the modular transformed functions of $1/t$, as appropriate in the small $t$ limit. Dividing by the spatial volume, and the circumference of Euclidean time, we have the following expression for the vacuum energy density of the type IB superstring with $N$ D9branes, thermal duality transformed after compactification on the twisted torus with $B$-field $|B_{09}|$ $=$ ${\rm tanh} ( \pi \alpha)$, and $\alpha $$=$$\beta_C/\beta$. A $T_9$-duality transformation likewise enables analysis of the short distance limit of the Dirichlet vacuum, with a stretched string extending along the interval $X^{\rm 9A}$, of length $R$$=$$R^9_{\rm IA}$. 
The argument of the $B$ field has asymptoted to its high temperature value, ${\rm tanh} (\pi \alpha) $$\to $$1$, and the high temperature asymptotic expansion of the Jacobi theta functions is in integer powers of $q$$=$$e^{-\pi/ t}$, exposing the $t$$\to$$0$ limit of the integrand: 
\begin{eqnarray}
F^{\rm (IA)}_{\rm ann}  =&& - N^2 V_9
  (8\pi^2 \alpha^{\prime} )^{-5}  \left ( 1 + {\rm tanh} ( \pi \alpha ) \right ) \int_0^{\infty}
{{dt }\over{t }} \cdot t^{-5 }  e^{-R^2 t /2 \pi \alpha^{\prime} }
\times \left [ \eta (it)^{-6 } \right ] \left [ {{ e^{-\pi  \alpha^2 t} \eta(it)}\over{ \Theta_{11} (  \alpha , it ) }}
\right ] \cr
&& \quad\quad\quad \quad \times \sum_{w_0 = -\infty}^{\infty} 
\exp \left [ -  {{4\pi^2 w_0^2 \beta_{\rm IA}^2  }\over{\alpha^{\prime }}} t \right ]  \cr
&&  \times
\left [ {{ \Theta_{0 0 } ( \alpha  , it )}\over{ e^{- \pi\alpha^2 t} \eta ( it ) }}
 \left ( {{ \Theta_{00} (0 , it )}\over{ \eta (it ) }} \right )^3
-
 {{ \Theta_{0 1 } ( \alpha/ , it )}\over{ e^{-\pi  \alpha^2 t} \eta (it ) }}
 \left ( {{ \Theta_{0 1} ( 0  , it )}\over{ \eta ( it ) }} \right )^3 \right ] 
 \cr && \cr
&&  \quad \quad + N^2 (8\pi^2 \alpha^{\prime} )^{-5} \left ( 1 + {\rm tanh} ( \pi \alpha ) \right )
\int_{0}^{\infty} {{dt}\over{t}} \cdot t^{-5} e^{-R^2 t /2 \pi \alpha^{\prime} } \times  \left [  [ \eta(it) ]^{-6} \right ] \cr
&& \quad \times {{1}\over{4}} 
 \left [ {{ e^{-\pi  \alpha^2 t} \eta(it)}\over{ \Theta_{11} (  \alpha  , it ) }}
\right ] \left [ 
 {{ \Theta_{1 0} (  \alpha  , it )}\over{ e^{- \pi  \alpha^2 t} \eta ( it  ) }}
 \left ( {{ \Theta_{10} (0, it )}\over{ \eta ( it ) }} \right )^3 + {{ \Theta_{11 } (  \alpha  , it )}\over{ e^{ - \pi  \alpha^2 t} \eta ( i t ) }}
 \left ( {{ \Theta_{11} (0 , i t )}\over{ \eta ( i t ) }} \right )^3 \right ]  \cr
&& \quad\quad\quad\quad  \times  \sum_{w_0= -\infty}^{\infty} 
\exp \left [ -  {{4\pi^2 w_0^2 \beta_{\rm IA}^2  }\over{\alpha^{\prime }}}  t \right ]  \quad ,
\label{eq:ibenj}
\end{eqnarray} 
where the last term from the R-R sector is only formal, since $\Theta_{11} (0,1t) $$=$$ 0$. 

\vskip 0.1in \noindent
Moving on to the corresponding results for the Mobius strip and Klein bottle, we express each worldsheet modular integral in terms of the variable $t$, where $t$ is the intrinsic length of either holes and crosscaps on the one-loop unoriented type IB string world sheets. For the Mobius strip topology, we have:
\begin{eqnarray}
{ \rho}^{\rm (IA)}_{\rm mob}  =&& - 2 N ( 2^5)
  (8\pi^2 \alpha^{\prime} )^{-5}  \left ( 1 + |B_{09}| \right ) \int_0^{\infty}
{{dt }\over{t }} \cdot t^{-5 }  e^{-R^2 t /2 \pi \alpha^{\prime} }
\times \left [  \eta (i t)^{-6 } \right ] \left [ {{ e^{-\pi  \alpha^2 t} \eta(i t)}\over{ \Theta_{11} (  \alpha , it ) }}
\right ] \cr
&& \quad\quad\quad \quad \times \sum_{w_0 = -\infty}^{\infty} 
\exp \left [ -  {{4\pi^2 w_0^2 \beta_{\rm IA}^2  }\over{\alpha^{\prime }}} t \right ]  \cr
&&  \times
\left [ {{ \Theta_{0 1 } ( \alpha  , it )}\over{ e^{- \pi\alpha^2 t} \eta ( it ) }}
 \left ( {{ \Theta_{01} (0 , it )}\over{ \eta (it ) }} \right )^3 
 {{ \Theta_{10 } ( \alpha/ , it )}\over{ e^{-\pi  \alpha^2 t} \eta (it ) }}
 \left ( {{ \Theta_{10} ( 0  , it )}\over{ \eta ( it ) }} \right )^3 \right ] 
 \cr && \quad\quad \quad  \quad \times  {{ e^{- \pi\alpha^2 t} \eta ( it ) }\over{ \Theta_{0 0 } ( \alpha  , it ) }}
 \left ( {{ \eta (it )}\over{\Theta_{00} (0 , it )  }} \right )^3  \quad ,
\label{eq:imobk}
\end{eqnarray} 
and likewise, summing unoriented type IB world sheets with the topology of a Klein bottle, we have:
\begin{eqnarray}
{ \rho}^{\rm (IA)}_{\rm kb}  =&& 2^{10}
  (8\pi^2 \alpha^{\prime} )^{-5}  \left ( 1 + |B_{09}| \right ) \int_0^{\infty}
{{dt }\over{t }} \cdot t^{-5 }  e^{-R^2 t /2 \pi \alpha^{\prime} }
\times \left [ \eta (it)^{-6 } \right ] \left [ {{ e^{-\pi  \alpha^2 t} \eta(i t)}\over{ \Theta_{11} (  \alpha , it ) }}
\right ] \cr
&& \quad\quad\quad \quad \times \sum_{w_0 = -\infty}^{\infty} 
\exp \left [ -  {{4\pi^2 w_0^2 \beta_{\rm IA}^2  }\over{\alpha^{\prime }}} t \right ]  \cr
&&  \times
\left [ {{ \Theta_{0 0 } ( \alpha  , it )}\over{ e^{- \pi\alpha^2 t} \eta ( it ) }}
 \left ( {{ \Theta_{00} (0 , it )}\over{ \eta (it ) }} \right )^3
-
 {{ \Theta_{0 1 } ( \alpha , it )}\over{ e^{-\pi  \alpha^2 t} \eta (it ) }}
 \left ( {{ \Theta_{0 1} ( 0  , it )}\over{ \eta ( it ) }} \right )^3 \right ] 
 \cr && \cr
&&  \quad \quad - 2^{10} (8\pi^2 \alpha^{\prime} )^{-5} \left ( 1 + |B_{09}| \right )
\int_{0}^{\infty} {{dt}\over{t}} \cdot t^{-5} e^{-R^2 t /2 \pi \alpha^{\prime} } \times  \left [   [ \eta(it) ]^{-6} \right ] \cr
&& \quad \times {{1}\over{4}} 
 \left [ {{ e^{-\pi  \alpha^2 t} \eta(it)}\over{ \Theta_{11} (  \alpha  , it ) }}
\right ] \left [ 
 {{ \Theta_{1 0} (  \alpha  , it )}\over{ e^{- \pi  \alpha^2 t} \eta ( it  ) }}
 \left ( {{ \Theta_{10} (0, it )}\over{ \eta ( it ) }} \right )^3 + {{ \Theta_{11 } (  \alpha  , it )}\over{ e^{ - \pi  \alpha^2 t} \eta ( i t ) }}
 \left ( {{ \Theta_{11} (0 , i t )}\over{ \eta ( i t ) }} \right )^3 \right ]  \cr
&& \quad\quad\quad\quad  \times  \sum_{w_0= -\infty}^{\infty} 
\exp \left [ -  {{4\pi^2 w_0^2 \beta_{\rm IA}^2  }\over{\alpha^{\prime }}}  t \right ]  \quad .
\label{eq:ikb}
\end{eqnarray}

\subsection{Low Temperature Massless Limit of Type I O(32) String}

\vskip 0.1in\noindent We begin with the low temperature limit of the annulus amplitude making a change of variable $y=$$A/t$ in order to make the integral representation of the Whittaker function evident:
\begin{eqnarray}
{ \rho}^{\rm (IB)}_{\rm ann}  =&& -
  (8\pi^2 \alpha^{\prime} )^{-5}  \left ( 1 + \pi \alpha \right ) \int_0^{\infty} 
{{dt }\over{t }} \cdot t^{-5 }   \sum_{m=0}^{\infty} \sum_{n_0= -\infty}^{\infty} \sum_{n_9= -\infty}^{\infty}  f_m^{\rm (IB)} (\alpha) \cr
&& \quad \quad \quad \quad \times ~ \exp{\left [ - \pi m t +  t \left ( {{  \alpha^{\prime} n_9^2}\over{R_{\rm IB}^2}} + {{4\pi^2 n_0^2 \alpha^{\prime }  }\over{\beta_{\rm IB}^2}} \right ) \right ]}
 \cr
 =&&  -
  (8\pi^2 \alpha^{\prime} )^{-5}  \left ( 1 + \pi \alpha \right ) \sum_{n_0= -\infty}^{\infty} \sum_{n_9= -\infty}^{\infty} \sum_{m=0}^{\infty}   f_m^{\rm (IB)} (\alpha) \cr
&&\quad \quad \times ~  \left [ m+ {{4\pi^2 n_0^2 \alpha^{\prime } }\over{\beta_{\rm IB}^2 }} + {{  \alpha^{\prime} n_9^2}\over{R_{\rm IB}^2}}  \right ]^{-5} \int dy y^4 e^{-1/y}  \cr
 =&&-  
  (8\pi^2 \alpha^{\prime} )^{-5}  \left ( 1 + \pi \alpha \right ) \sum_{n_0= -\infty}^{\infty} \sum_{n_9= -\infty}^{\infty} \sum_{m=0}^{\infty}   f_m^{\rm (IB)} (\alpha) \cr
&&\quad \quad \times ~  \left [ m+ {{4\pi^2 n_0^2 \alpha^{\prime } }\over{\beta_{\rm IB}^2 }} + {{  \alpha^{\prime} n_9^2}\over{R_{\rm IB}^2}}  \right ]^{-5} A^2 \Gamma (-5) e^{A/2} 
{\cal W}_{3,-5/2} (A) \quad . 
\quad , 
\label{eq:iffnn}
\end{eqnarray}
Substituting in this expression the power series expansion of the Whittaker function gives the result:
\begin{eqnarray}
{ \rho}^{\rm (IB)}_{\rm ann}  =&& -
 (8\pi^2 \alpha^{\prime} )^{-5}  \left ( 1 + \pi \alpha \right ) \sum_{n_0= -\infty}^{\infty} \sum_{n_9= -\infty}^{\infty} \sum_{m=0}^{\infty}   f_m^{\rm (IB)} (\alpha) 
 \left [ m+ {{4\pi^2 n_0^2 \alpha^{\prime } }\over{\beta_{\rm IB}^2 }} + {{  \alpha^{\prime} n_9^2}\over{R_{\rm IB}^2}}  \right ]^{-5} \cr 
&&\quad \quad \times ~ {{ \Gamma (-5) (-1)^5}\over{\Gamma (1) \Gamma (-5)}}  \{ \sum_{k=0}^{\infty} {{\Gamma (k-5) }\over{k! (k-5)!}}  A^k \left ( \psi (k+1)+ \psi (k-4) - \psi (k-5) 
- {\rm ln} (A) \right )  \cr
&& \quad \quad \quad \quad \quad \quad + ~ (-A)^5 \sum_{k=0}^4 {{\Gamma (5+k) \Gamma (k)}\over{k!}} (-A)^k  \} \quad .
\label{eq:iffnnh}
\end{eqnarray}
Expanding about the massless limit, and setting the degeneracies to the massless bosonic spacetime modes alone, we can sum the thermal momentum modes to extract the zeta function, $\zeta (-2 , 0)$, and the $T^{10}$ leading behavior of the low energy finite temperature gauge theory:
\begin{eqnarray}
{ \rho}^{\rm (IB)}_{\rm ann}  \simeq&& -
 (8\pi^2 \alpha^{\prime} )^{-5}  \sum_{n_0= -\infty}^{\infty} \sum_{n_9= -\infty}^{\infty}  b_0^{\rm (IB)}  
 \left [  {{4\pi^2 n_0^2 \alpha^{\prime } }\over{\beta_{\rm IB}^2 }} + {{  \alpha^{\prime} n_9^2}\over{R_{\rm IB}^2}}  \right ]^{-5} \cr 
&&\quad \quad \times ~ {{ \Gamma (-5) (-1)^5}\over{\Gamma (1) \Gamma (-5)}}  \{ \sum_{k=0}^{\infty} {{\Gamma (k-5) }\over{k! (k-5)!}}  A^k \left ( \psi (k+1)+ \psi (k-4) - \psi (k-5) 
- {\rm ln} (A) \right )  \cr
&& \quad \quad \quad \quad \quad \quad + ~ (-A)^5 \sum_{k=0}^4 {{\Gamma (5+k) \Gamma (k)}\over{k!}} (-A)^k  \} \cr  \simeq&& -
 (8\pi^2 \alpha^{\prime} )^{-5}  T^{10} \zeta (-2,0) ~ 496    (4\pi^2 \alpha^{\prime } )^{-5}
 \left \{ 1 \right \} \quad .
\label{eq:iffnnj}
\end{eqnarray}

\vskip 0.1in\noindent Finally, we note that the normalization of the heterotic ${\rm Spin ({\rm 32})}/{\rm Z}_2$ string vacuum energy density can be determined from this result by matching with the corresponding graph of the Type IB superstring, since the massless 496 gauge bosons of the spacetime gauge group are restricted to the oriented open string sector. Comparing with the analogous zero temperature spacetime bosonic massless mode limit of the heterotic string one loop amplitude:
\begin{eqnarray}
\rho_{\rm H} (\beta) \simeq  && - 2 {\cal N} 496  ~ {{1}\over{5!}} ~ (4\pi^2 \alpha^{\prime})^{-5} \times 
\quarter \left (  4\pi^2 \alpha^{\prime} \right )^5 T^{10}       \quad ,
\label{eq:masszerok}
 \end{eqnarray}
we find the simple result:
\begin{equation}
2^{-5}  = 2 {\cal N} {{1}\over{5!}} \quarter \quad .
\label{eq:norm}
\end{equation}

\subsection{High Mass Level Limit of Type IB ${\rm Spin ({\rm 32})}/{\rm Z}_2$ String}

\vskip 0.1in\noindent We begin with the high temperature limit of the one loop vacuum energy density of the open oriented sector derived above, recognizing in that expression
the integral representation of the modified Bessel function:
\begin{eqnarray}
{ \rho}^{\rm (IB)}_{\rm ann}  =&& 
  (8\pi^2 \alpha^{\prime} )^{-5}  \left ( 1 + {\rm tanh} (\pi \alpha ) \right )  \int_0^{\infty} 
{{dt }\over{t }} \cdot t^{-5 }   \sum_{m=0}^{\infty} \sum_{w_0= -\infty}^{\infty} \sum_{n_9= -\infty}^{\infty}  \cr 
&& \quad \quad \quad \times ~ f_m^{\rm (IB)} (\alpha)\exp{\left [ - {{\pi m}\over{t}}  - t \left ( {{  \alpha^{\prime} n_9^2}\over{R_{\rm IB}^2}} + {{4\pi^2 w_0^2 \beta_{\rm IB}^2  }\over{\alpha^{\prime }}} \right ) \right ]}
 \cr
 \simeq&&  
  (8\pi^2 \alpha^{\prime} )^{-5} \left ( 1 + {\rm tanh} (\pi \alpha) \right ) \sum_{m=0}^{\infty} f_m^{\rm (IB)} (\alpha) \sum_{w_0= -\infty}^{\infty} \sum_{n_9= -\infty}^{\infty}  \cr
&&\quad \quad \times ~  (m\pi)^{5/2} \left [  
 {{4\pi^2 w_0^2 \beta_{\rm IB}^2  }\over{\alpha^{\prime }}} + {{  \alpha^{\prime} n_9^2}\over{R_{\rm IB}^2}}   \right ]^{-5/2}  K_{5} (z)   \quad . 
\label{eq:ikbnj}
\end{eqnarray}
The Bessel function can be replaced by its asymptotic expansion (GR 8.446.1) in the limit of high mass level numbers, also setting the tanh function to unity, and 
the $f_m$ to their values at the self dual temperature:
\begin{eqnarray}
{ \rho}^{\rm (IB)}_{\rm ann}  =&& 
   2(8\pi^2 \alpha^{\prime} )^{-5} \sum_{m=0}^{\infty} f_m^{\rm (IB)} (1) \sum_{w_0= -\infty}^{\infty} \sum_{n_9= -\infty}^{\infty}  \cr
&&\quad \quad \times ~  (m\pi)^{5/2} \left [  {{4\pi^2 w_0^2 \beta_{\rm IB}^2  }\over{\alpha^{\prime }}} + {{  \alpha^{\prime} n_9^2}\over{R_{\rm IB}^2}}   \right ]^{-5/2} K_{5} (z)   \quad . 
\label{eq:ikbnp}
\end{eqnarray}
Restricting to the degeneracies of the bosonic spacetime modes alone, $b_m^{\rm (IB)} $, the result is a damping of the Hagedorn growth of the numerical degeneracies for large level number by the exponential of the square root of the mass level number with a coefficient which is always large at high mass level numbers and high temperature. It is helpful to T-dualize to the thermal dual large radius, $\beta_{\rm IA}$, since the thermal IB coordinate is approaching small radius:
\begin{eqnarray}
{ \rho}^{\rm (IB)}_{\rm ann}  =&& 
   2(8\pi^2 \alpha^{\prime} )^{-5} \sum_{m=0}^{\infty} b_m^{\rm (IB)} (1) \sum_{w_0= -\infty}^{\infty} \sum_{n_9= -\infty}^{\infty}  \cr
&&\quad \quad \times ~  (m\pi)^{5/2} \left [  {{4\pi^2 w_0^2 \beta_{\rm IB}^2  }\over{\alpha^{\prime }}} + {{  \alpha^{\prime} n_9^2}\over{R_{\rm IB}^2}}   \right ]^{-5/2}  \cr
\simeq && 2(8\pi^2 \alpha^{\prime} )^{-5} \sum_{m=0}^{\infty} b_m^{\rm (IB)} (1) \sum_{w_0= -\infty}^{\infty} \sum_{n_9= -\infty}^{\infty}  \cr
&& \quad \times  ~ m^{9/4} 2^6 \pi^{-5/2} e^{- 4\pi  w_0 T_{IA} \alpha^{\prime 1/2}   
                 [  1 + {{  n_9^2 \beta_{\rm IA}^2}\over{w_0^2 R_{IB}^2 }} ]^{1/2} {\sqrt m}  }  \quad . 
\label{eq:ikbnd}
\end{eqnarray}
This completes our demonstration of the finiteness of the Type IB open and closed superstring theory. It should be noted that the high mass level number limit of the unoriented graphs 
are also integral representations of the modified Bessel function, and their asymptotic growth can be analyzed similarly.

\section{Type I Pair Correlator of Spacelike Wilson Loops}

\vskip 0.1in \noindent 
In the massless mode, field theoretic, limit of the Type IB superstring amplitude annulus graph, the spacelike Wilson loop expectation value \cite{dkps,pairb,pairf,ncom} is the change in the internal energy of the finite temperature gauge theory vacuum due to the introduction of an infinitely massive quark in the presence of the external NS two-form field.\footnote{A shift of the NS twoform potential by an external abelian gauge field strength gives the result in the presence of an external constant chromoelectric field, with slow-moving heavy quarks, or, in an external chromomagnetic field, with static heavy quarks \cite{dkps,pairb,pairf}.} The spacelike Wilson loop is the world history of a semiclassical heavy charged color source living in the fundamental representation of an $O(4)$ subset of the $O(16)$ gauge group. The coincidence of two D8branes, and their orientifold images, gives 4 additional massless, zero length, open string modes, states completing the ${\bf 3 \oplus \bf 3}$ of the $O(4)$$\simeq$$SU(2)$$\times$$SU(2)$ gauge group. Thus, the single spacelike Polyakov-Susskind loop at spatial coincidence is the world-history of a heavy quark in the $\bf 2 \oplus 2$ representation of $O(4)$. Namely, the parallel stack of 2 D8branes, and their two orientifold image D8branes, at one of the orientifold planes of the O(16)$\times$O(16) type IA string compactified on a twisted torus, coincide to give all of the massless zero length open strings in the adjoint representation of $O(4)$, and the Chan-Paton factor for the endpoints themselves, i.e., the end-point wave function, transforms in the ${\bf 2 \oplus 2}$ fundamental irrep of $SU(2)$$\times$$SU(2)$. Note that the Wilson loop operator always contains a trace over the representation of the nonabelian gauge group. 

\vskip 0.1in  \noindent 
The pair correlator of spacelike Polyakov-Susskind loops, ${\cal W}^{(2)}$, can be derived from first principles. Incorporating the changes required by finite temperature for the Type IB superstring compactified on the twisted torus, and using the results of \cite{pairb,pairf} for superstring amplitudes with macroscopic incoming and outgoing strings, gives the following expression for the open oriented contribution to the pair correlator of parallel spacelike loops spatially separated by a distance $R_{\rm IB}$:
\begin{eqnarray}
{\cal W}_{\rm IB}^{(2)}  =&&
\left ( 1 + |{\rm tanh} (\pi\alpha)| \right ) \int_0^{\infty} {{dt}\over{2t}}  ~ (2t)^{1/2} {{e^{- R_{\rm IB}^2 t/2\pi \alpha^{\prime}
}}\over{\eta(it)^{6}}} \left [ {{ e^{i\pi t \alpha^2} \eta(it)}\over{ \Theta_{11} (  \alpha , it ) }}
\right ] \cr
&& \quad\quad\quad \quad \times \sum_{n_i = -\infty}^{\infty} 
\exp \left [ - \pi \left (  {{  \alpha^{\prime} n_9^2}\over{R_{\rm IB}^2}}  + {{ 4 \pi^2 \alpha^{\prime} n_0^2}\over{\beta_{\rm IB}^2}} \right ) t \right ] \cr
&&  \quad \times
\left [ {{ \Theta_{0 0 } ( \alpha , it )}\over{ e^{i\pi t \alpha^2} \eta ( it ) }}
 \left ( {{ \Theta_{00} (0 , it )}\over{ \eta (it ) }} \right )^3
-
 {{ \Theta_{0 1 } ( \alpha , it )}\over{ e^{i\pi t \alpha^2} \eta (it ) }}
 \left ( {{ \Theta_{0 1} ( 0  , it )}\over{ \eta ( it ) }} \right )^3
-
 {{ \Theta_{1 0} (  \alpha , \tau )}\over{ e^{i \pi t \alpha^2} \eta ( it ) }}
 \left ( {{ \Theta_{10} (0, it )}\over{ \eta ( it ) }} \right )^3
\right ]
 \cr && \cr
&&  \quad \quad 
- \left ( 1 + |{\rm tanh} (\pi \alpha) | \right )
\int_{0}^{\infty} {{dt}\over{2t}} \cdot (2t)^{1/2} [ \eta(it) ]^{-6} e^{- R_{\rm IB}^2 t/2\pi \alpha^{\prime}} \cr
&& \quad \quad \quad \quad\quad \quad \times {{1}\over{4}} 
 \left [ {{ e^{i\pi t \alpha^2} \eta(it)}\over{ \Theta_{11} (  \alpha , it ) }}
\right ]  \left [ {{ \Theta_{11 } (  \alpha , it )}\over{ e^{ i \pi t \alpha^2} \eta ( it ) }}
 \left ( {{ \Theta_{11} (0 , it )}\over{ \eta ( it ) }} \right )^3 \right ]  \cr
&& \quad\quad\quad\quad  \times  \sum_{n_i = -\infty}^{\infty} 
\exp \left [ - \pi \left (   {{ \alpha^{\prime} n_9^2}\over{R_{\rm IB}^2}} +  {{ 4\pi^2 \alpha^{\prime} n_0^2}\over{\beta_{\rm IB}^2}}  \right )  t  \right ]
\label{eq:iben}
\end{eqnarray} 
Note that the expression above is valid for all values of $T=1/\beta_{\rm IB}$, with $B$ and $\beta$ both target spacetime moduli, with $|B|$ linear for small $T$, and asymptoting to unity at temperatures approaching the string deconfinement scale. 

\vskip 0.1in \noindent
The massless Yang-Mills gauge field theory limit of ${\cal W}_{\rm IB}^{(2)}$  yields the potential between two heavy color charged sources at spatial separations $R_{\rm IB}$ $>$ $\alpha^{\prime 1/2}$. We work in the large radius limit, and at type IB temperatures much below the string scale. Note that the amplitude ${\cal W}^{(2)}$ is dimensionless. We will find that this is a good paradigm for a nonabelian gauge theory in the {\em deconfinement} regime; all of the thermal excitations are Matsubara modes, namely, thermal momenta, and the type IB superstring has no winding modes. The heavy quark pair can be pulled out to spatial separations larger than the string scale, giving clear evidence for the inverse linear term which is universal--- the Luscher term, common to all effective Nambu-Goto-Eguchi-Schild-Polyakov QCD strings \cite{effective,luscher,newlw}. In addition, we can also derived the systematic thermal corrections to the Luscher potential. 

\vskip 0.1in \noindent
Retaining the leading terms in the $q$ expansion, dominated by thermal momentum modes at low temperatures far below the string mass scale, and performing an explicit term-by-term integration over the world-sheet modulus, $t$, isolates the leading terms in the massless string spectrum, $m=0$, and thermal and spatial Kaluza-Klein modes. In the low temperature regime, the inverse temperature lies within the range, $ 2\pi \alpha^{\prime 1/2} << R_{\rm IB} << \beta_{\rm IB} $, in string scale units \cite{pairf}, and we substitute the power law expansion for the gamma function, after a change of variable, $t$$\to$$At$. The argument of the gamma function, $\Gamma (z)$, takes the form, $|z|<1$:
\begin{equation}
A =  \left [ m+  {{  n_9^2 \alpha^{\prime} }\over{R_{\rm IB}^2}}  + {{ R_{\rm IB}^2 }\over{ 4 \pi^2 \alpha^{\prime} }} + {{n_0^2  4 \pi^2 \alpha^{\prime}  }\over{ \beta_{\rm IB}^2}}  \right ]  \quad  ,
\label{eq:expon}
\end{equation} 
\vskip 0.2in \noindent
expanding about $m$$=$$0$. The result of the modular integral can be expressed in terms of the power series expansion of the gamma function, with argument $z$$=\half$:
\begin{eqnarray}
{\Gamma} (z+1) = &&  \sum_{k=0}^{\infty} c_k z^k ~, \quad c_0 = 1, ~ c_1 = - {\rm C}, ~ c_{n+1} = \sum_{k=0}^n {{(-1)^{k+1} s_{k+1} c_{n-k}}\over{n+1}}  , \quad 
s_1 = {\rm C} , ~ s_{n} = \zeta (n)  . \cr
&& 
\label{eq:whtt}
\end{eqnarray}
which gives the result:
\begin{eqnarray}
{\cal W}_{\rm IA}^{(2)}  =&&
2^{-1/2} \left ( 1 + |{\rm tanh} (\pi\alpha)| \right )  \sum_{m=0}^{\infty} f_{m}^{\rm (IB)} (\alpha) \sum_{n_0= -\infty}^{\infty} \sum_{n_9 = -\infty }^{\infty} \int_0^{\infty} {{dt}\over{t}}  ~ t^{1/2} e^{-t}  \cr && \cr
&& \quad\quad\quad\quad  \times  \left ( {{R^2}\over{ 2 \pi^2  \alpha^{\prime }}}   + {{  \alpha^{\prime} n_9^2}\over{R_{\rm IB}^2}}  + {{n_0^2 4 \pi^2 \alpha^{\prime}  }\over{ \beta_{\rm IB}^2 }}  \right )^{-1/2}  
  \cr 
\simeq&&   2^{-1/2}  \Gamma (1/2)  \left ( 1 + {\rm tanh} (\pi\alpha) \right ) f_0^{\rm IB} (\alpha) \times   \left ( {{R^2}\over{ 2 \pi^2  \alpha^{\prime }}}   + {{  \alpha^{\prime} n_9^2}\over{R_{\rm IB}^2}}  + {{n_0^2 4 \pi^2 \alpha^{\prime}  }\over{ \beta_{\rm IB}^2 }}          \right )^{-1/2}  \cr
\simeq&& 2^{-1/2}  \Gamma (1/2)  ( 1 + \pi\alpha )  \times   \left ( {{R^2}\over{ 2 \pi^2  \alpha^{\prime }}}   + {{  \alpha^{\prime} n_9^2}\over{R_{\rm IB}^2}}  + {{n_0^2 4 \pi^2 \alpha^{\prime}  }\over{ \beta_{\rm IB}^2 }}          \right )^{-1/2} \cr && \cr 
&& \quad \quad \times \left [  2(2 {\rm Cosh} (2[\pi {\rm tanh} (\pi\alpha) ] )  + 6 ) - 16 {\rm Cosh} ([\pi {\rm tanh} (\pi \alpha) ]) \right ] \cr && \cr 
=&& \Gamma (1/2) ( 1 + \pi\alpha ) \left [  16 - 16 (1+ (\pi \alpha)^2 ) \right ] \cr
 && \quad \quad\quad \times  \left [   {{ \pi \alpha^{\prime 1/2}}\over{R_{\rm IB} }}  -  \zeta (-2 , 0 ) \left ( {{4 \pi^5 \alpha^{\prime 5/2} }\over{\beta_{\rm IB}^2 }} \right ) {{1}\over{R_{\rm IB} ^3}}  + O \left ( \alpha^{\prime 9/2}/R_{\rm IB}^5 \beta_{\rm IB}^4 \right)  \right ]    \quad ,
\label{eq:static}
\end{eqnarray}
where we recall that the inverse temperature lies within the range, $ 2\pi \alpha^{\prime 1/2} < \beta_{\rm IB} << R_{\rm IB} $, in string scale units \cite{pairf}.  Our result shows that the leading correction to the inverse linear attractive potential, namely, the universal Luscher term, in the zero temperature static heavy quark potential, is $O(1/R^3)$, taking the form of a systematic series expansion in powers of $(\alpha^{\prime 2 }/ \beta_{\rm IB}^2 R^2)$ at type IB temperatures far above the thermal duality transformation temperature, namely, the string mass scale, $T_C$ $=$ ${{1}\over{2\pi}} \alpha^{\prime -1/2}$.\footnote{Note the remarks by Luscher and Weisz in \cite{newlw} on the absence of a $1/R^2$ term in the heavy quark potential, presciently arguing in favor of the leading $1/R^3 $ correction. It should be noted their argument is valid on general grounds, and is not specific to the finite temperature gauge theory, but it is exactly what we too find in our derivation from string theory of the static heavy quark potential at finite temperature.} 

\vskip 0.1in \noindent We now perform both a thermal duality transformation on the expression for the Type IB pair correlator of spacelike Wilson loops, in addition to a spatial 
$T_9$-duality transformation. Expressing the result in terms of Type IA string variables, the target spacetime geometry is that of a stack of 32 thermal D8branes in the 10D Type IA O(32) superstring compactified on $R^8$, with an $S^1/{\rm Z}_2$$\times$$S^1/{\rm Z}_2$ orthogonal to the worldvolume of the thermal D8brane stack.\footnote{A thermal Dpbrane has only $p$ noncompact coordinates, and a p-dimensional Euclidean, spatial, worldvolume. The gauge fields supported on this brane are finite temperature supersymmetric gauge theories in $(p+1)$- dimensions.} We consider a pair of heavy colored sources whose world-histories are loops winding along $X^0_{\rm E}$, which is now an interval of length $\beta_{\rm IA}$; hence the world histories of the infinitely heavy \lq\lq quarks" are stretched parallel to the Euclidean time interval. In addition, they are spatially separated by a Dirichlet-string of length $R$, stretched parallel to the interval $X_9^{\prime}$. Note that upon T-dualizing both the $X^9$ and $X^0$ coordinates, we obtain a Type IA superstring theory, with a tower of spatial, and thermal, winding modes, replacing the thermal momentum modes of the Type IB superstring. Thus, the infinitely heavy color sources are now confined in a bound state with spatial separation within a Type IA string length; remarkably, we will show that we can nevertheless derive analytical expressions for the binding energy. 

\vskip 0.1in \noindent
We find that this novel type TA phase is a good model for the {\em confinement} phase of nonabelian gauge theories, with its tower of thermal and spatial winding string modes. Comparing with the expression for the pair correlator of spacelike Wilson loops in the finite temperature type IB superstring theory given in Eq.\ (4), we can repeat the steps taken above, and extract the massless level $m$ $=$ $0$ low energy gauge theory limit. The result is:
\begin{eqnarray}
{\cal W}^{(2)}_{\rm Linear} \simeq&&   \alpha^{\prime -1/2} \Gamma (1/2) \left [  2(2 {\rm Cosh} (2[\pi {\rm tanh} (\pi \alpha) ] )  + 6 ) - 16 {\rm Cosh} ([\pi {\rm tanh} (\pi \alpha) ]) \right ]   \cr && \cr
 && \quad \times   R \left [ 1  - {{1}\over{2}} \sum_{w_0=-\infty}^{\infty} (   {{4\pi^2 }\over{\alpha^{\prime 2}}} \left (  w_0^2 \beta_{\rm IA}^2 R^2  \right ) \right ]  \cr 
&& \cr  =&&
 \alpha^{\prime -1/2} \Gamma (1/2) \left [  2(2 {\rm Cosh} (2[\pi {\rm tanh} (\pi \alpha) ] )  + 6 ) - 16 {\rm Cosh} ([\pi {\rm tanh} (\pi \alpha) ]) \right ]  \cr && \cr 
&& \quad \times  R  \left [ 1  - \zeta (-2, 0 ) \alpha^{\prime -2} \beta_{\rm IA}^2 R^2 \right ] \quad .
\label{eq:statici}
\end{eqnarray}
This expression gives the subleading thermal correction to the linear potential for a pair of semiclassical heavy quarks, with spatial separation $R$, and with a Dirichlet spatial winding mode string stretched between them. Pulling apart the heavy colored sources, gives a potential that grows linearly with $R$ and the Dirichlet string is the confining string. There is a potential energy cost to separating the color charged sources, and the temperature dependent term is a further suppression, pointing to confinement.

\vskip 0.1in \noindent
Comparing with Eq (17), note that the binding energy density is a continuous function of temperature at the string scale critical temperature, but the first derivative with respect to temperature has a discontinuity. Remarkably, precise computations can nevertheless be carried out on either side of the phase boundary in temperature by, respectively, working in the respective low energy gauge theory limits of thermal and spatial dual string theories, Type IB and Type IA. Thus, our results are clearly suggestive of a thermal deconfinement phase {\em transition} at $T_C$ $=$ $1/2\pi\alpha^{\prime 1/2}$ in the type IA gauge theory, and this deconfining phase transition can be identified as {\em first order}. Most pertinent, note that taking $R$ $\to$ $0$ smoothly gives a vanishing expectation value for the single Polyakov-Susskind loop in the confinement regime, as was conjectured for the order parameter of the thermal deconfinement transition--- for the O(2n) groups the center symmetry is ${\rm Z}_2 $$\times$$ {\rm Z}_2$. This completes our discussion of the thermal deconfinement transition and its order parameter for the O(32-2n) anomaly free nonabelian gauge theory limit of the finite temperature Type I superstring. In closing, we should point out that we have barely touched the considerable information in the full string pair correlation function, and the thermal spectrum with massive winding modes, which remains for future analysis.

\section{Conclusions} 

\vskip 0.1in \noindent
The original suggestion that the Polyakov string path integral might provide a renormalizable analytical description of the expectation value of a Wilson loop valid to arbitrarily short distances was made by Orlando Alvarez in \cite{alvarez}, although its implementation at the time was stymied by many technical, and conceptual, problems. The extension of the string path integral formalism for on-shell scattering amplitudes to those for the off-shell closed string tree propagator, incorporating the modified Dirichlet, or Wilson loop, boundary conditions proposed by Alvarez in \cite{alvarez}, was given by Cohen, Moore, Nelson, and Polchinski \cite{cmnp}. The suggestive sketch of the computation given in \cite{cmnp} was subsequently reformulated by me in collaboration with my students Yujun Chen and Eric Novak \cite{pairb,pairf,ncom}, giving a proper implementation of the super boundary reparametrization invariance, that was also Weyl invariant, and for macroscopic Wilson loops. We incorporated also the modern framework of Dirichlet strings, and Dbranes and orientifold planes in background twoform field strengths \cite{dkps} \cite{ncom,flux}. Most importantly, a discussion of thermal deconfinement required development of a consistent Euclidean time quantization of finite temperature superstring theory, settling also the troubling issue of the Hagedorn divergence of the degeneracies of the string mass level expansion, which is suppressed by a compensating term arising from the integral over worldsheet moduli that preserves worldsheet reparametrization invariance.

\vskip 0.1in \noindent
In summary, we have provided in this paper a precise answer to the classic question raised originally by Kenneth G.\ Wilson \cite{willat}, and by Peskin and Alvarez \cite{peskin,alvarez}, re-posed most recently by Luscher and Weisz \cite{luscher,newlw}: {\em Are the lattice gauge theory results suggestive of the existence of a {\em renormalizable alternative to the Nambu-Goto effective string model} which can account for the unusually good accuracy of effective string models for QCD flux tubes of finite thickness extending into the short distance regime, far beyond what was a reasonable expectation of an effective string model description of confinement?} Luscher and Weisz, in particular, in recent papers \cite{newlw} have brought to light the unusual accuracy of effective string models of fluxtubes linking heavy quark pairs--- even in the short distance regime--- analytical results that provide valuable corroboration of lattice gauge theory simulations, as also comprehensively examined by Juge, Kuti, and Morningstar \cite{kuti}. 

\vskip 0.1in \noindent
Our answer to the question is, of course, yes, and, furthermore, that the Polyakov macroscopic string path integral framework \cite{cmnp,pairb,pairf} provides not merely an effective QCD string model for the computation of the expectation value of the spacelike history of a semiclassical heavy quark, but gives strong evidence for the veracity of 
heterotic-type I superstring theories as accurate descriptions of the real world at particle accelerator scale short distances and high energies, with a precise derivation of their low energy gauge theory limit. In particular, our work goes beyond the original analyses by Brink, Green, and Schwarz, and others \cite{bgs,schwarz,backg}, by performing in closed form the integrals over worldsheet moduli, and thereby preserving the worldsheet super Diff$\times$Weyl gauge symmetries, when taking both the low energy supergravity and nonabelian gauge theory limits \cite{pairb,pairf}, in the presence of background two form field strengths \cite{ncom}, and at finite temperature. Thus far, our considerations have been restricted to one-loop string amplitudes, and it is fascinating to observe the wealth of new physics which can be extracted with the inclusion of macroscopic superstring amplitudes. In particular, it has been interesting to compare our results with the analytic gauge theory methodology pursued by Erich Poppitz et al \cite{poppitz}, and we look forward to the further interface of analytic and lattice gauge theory with both string perturbation theory, including string tree, and two loop, results, and the non-perturbative estimates of the super potential and scalar potential in the SYM* theory of \cite{poppitz}, namely, for the low energy softly broken supersymmetry breaking phenomenon in the heterotic/type I superstrings.  

\vskip 0.1in \noindent Finally, in a series of ground-breaking works \cite{witn}, in part with Donagi, Witten has convincingly argued that there is no ambiguity in the measure in the type IIB(IIA) and heterotic superstring path integrals at higher genus, and in on-shell superstring scattering amplitudes, arising from the required integrals over supermoduli. This analysis has also been extended to superstring multiloop amplitudes with Ramond-Ramond punctures. Compactifications of the Type IB (IA) superstring on $K3 \times T^2/{\rm Z}_2$ would result in 4d chiral N=1 spacetime supergravity-super Yang Mills fields, with target spacetime fermions living in the spinor representations of the chiral subgroups of $E_8$$\times$$E_8$, built up from the $O(16)$$\times$$O(16)$ Type IA (IB) theory. Hence, the crucial phenomenologically interesting vacua resembling the Standard Model quarks and leptons in addition to the gauge fields, appear deserving of further development of the worldsheet techniques for target spacetime fermion on-shell, and macroscopic, string scattering amplitudes, and should lead to further comparisons with both analytic, and numerical lattice gauge theory, investigations.

\vskip 0.2in \noindent
{\large{\bf Acknowledgments:}} I would like to acknowledge Paul Plassmann for his unstinting encouragement and patience bringing this work to completion. I would like to acknowledge my collaborators, Yujun Chen and Eric Novak, for their helpful participation in the early stages of this research. I thank Erich Poppitz for his interest, and useful comments on the manuscript. Paul Aspinwall, David Gross, Andreas Kronfeld, Juan Maldacena, Stephen Shenker, Herman Verlinde, and Edward Witten have been helpful at various stages during this research, which was brought to completion in the stimulating environs of the Lewis-Fine and Firestone Libraries at Princeton University.

\vskip 0.3in \noindent 
\section{Appendix: Aspects of Heterotic String Theory}

\vskip 0.1in \noindent In this appendix, we clarify certain aspects of the worldsheet formalism of the heterotic strings, showing their intimate relation to the basic building blocks, the Type IIB and the Type IIA superstrings, and which sheds light on the strong-weak dualities of the superstrings and M theory. 

\vskip 0.3in \noindent
\subsection{Chiral $N_S =1$ Orbifolds of the 10D Type II
Superstrings}
 
\vskip 0.1in \noindent 
Recall that the Type II superstring theories are named
by the parity of their 32 component 10D Majorana-Weyl 
spinors. In the Type IIA string theory, the 10D
spinors, and massless spin 3/2 gravitinos,
have opposite spacetime parity, whereas in the chiral
IIB string theory, they have identical spacetime
parity. This property distinguishes the two type II
superstrings. Recall that spacetime parity is given by
the product of left and right worldsheet parities.
We will show that the Hilbert space of the type IIA 
superstring admits {\em two} inequivalent $N_S$ = 1
chiral projections, where the subscript denotes the 10D 
target spacetime
supersymmetry. One of which eliminates the
higher rank pform potentials of the Ramond-Ramond
sector, whereas the other gives the T$_9$-dual of the
familiar type IB orientifold. 

\vskip 0.1in \noindent
We will follow the pedagogical derivation of the sum
over spin structures for the type IIA and type IIB
superstrings,
given in Section 10.6 of \cite{polchinskibook}. Each
has two equivalent, and self-consistent, choices of
sum over spin structures, which result in the GSO
projection to states with even spacetime G-parity,
respectively, chiral and non-chiral ten-dimensional
type II superstrings. We begin with the $N_S =1 $ chiral
projection of the Hilbert space of the type IIIB
superstring leading
to the well-known type IB orientifold
\cite{schwarz}:
\begin{eqnarray}
{\rm type ~ IIB}: \quad  &&((NS+,NS-) \oplus (NS-,
NS+))_s ,
(R-,NS+), (NS+,R-),  (R-,R-) \cr
{\rm type ~ IB}: \quad && ((NS+,NS-) \oplus (NS- ,
NS+))_s 
, ((NS+,R-) \oplus
(R-,NS+))_s,  (R-,R-) \quad . \cr
&&
\label{eq:specb}
\end{eqnarray}
Notice that the alternative choice of  Hilbert space
and worldsheet parity assignments in 
\cite{polchinskibook} gives the same result with an
$\Omega$ projection, since the type IIB superstring is
a chiral theory:
\begin{eqnarray}
{\rm type ~ IIB}^{\prime}: \quad  &&(NS+,NS+),
(R+,NS+),
(NS+,R+),  (R+,R+) \cr
{\rm type ~ IB}^{\prime} : \quad && (NS+,NS+),
((NS+,R+) \oplus
(R+,NS+))_s,  (R+,R+) \quad .
\label{eq:specbc}
\end{eqnarray}
Note that the 10d vector spinor now has positive
spacetime parity, but the theory is identical
to that above, a mere rewriting of the type IB
orientifold projection.

\vskip 0.1in \noindent Let us now contrast this with
the inequivalent $N_S =1$ chiral projections of the 10d
non-chiral Type IIA superstring, which can be written
as, see
Chapter 10.6 of \cite{polchinskibook}:
\begin{eqnarray}
{\rm type ~ IIA}: \quad  &&(NS+,NS+), (R+,NS+),
(NS+,R-),  (R+,R-) \cr
{\rm type ~ IIA'}: \quad && (NS+,NS+), (NS+,R+),
(R-,NS+), (R-,R+) \quad .
\label{eq:spec}
\end{eqnarray}
Notice that we have the freedom to symmetrize the type
IIA superstring Hilbert space over both choices of
GSO convention for ${\tilde{F}}$: $e^{\pi i F}$$=$$1$;
with $e^{\pi i {\tilde{F}} }$$=$$+1(R)$, $-1(NS)$, and
$e^{\pi i {\tilde{F}} }$$=$$-1(R)$, $+1(NS)$:
\begin{eqnarray}
{\rm type ~ IIA} :  &&  ((NS+,NS-) \oplus (NS- ,
NS+))_s, 
((NS+,R+) \oplus (R+,NS+))_s , \cr 
&& \quad ((R-,NS+)  \oplus  (NS+,R-) )_s, 
((R-,R+) \oplus (R+,R-))_s \cr
{\rm type ~ IIA}^{\prime}:  &&  (NS+,NS+) , 
((NS+,R+) \oplus (R+,NS+))_s , \cr 
&& \quad ((R-,NS+)  \oplus  (NS+,R-) )_s, 
((R-,R+) \oplus (R+,R-))_s \quad .
\label{eq:specs}
\end{eqnarray}
Under an $\Omega$ projection on the former, only
states symmetric under the interchange of left and
right movers remain, which eliminates one of the 10D
vector spinors, giving
the 10d N=1 type I$^{\prime}$, or type IA, string:
 \begin{equation}
{\rm type ~ IA} :   ((NS+,NS-) \oplus (NS- ,
NS+))_s , 
((R- , NS+) \oplus (NS+,R-))_s , ((R-,R+) \oplus
(R+,R-))_s .
\label{eq:spec1}
\end{equation}
Note that the left and right worldsheet parity of the
states in the Virasoro tower in which belongs the 10D
spacetime vector-spinor are {\em opposite}, giving a
spacetime spinor with {\em negative} 10D parity.

\vskip 0.1in \noindent
The Hilbert space of the type IIA superstring 
allows an even simpler $N_S $ chiral truncation
which eliminates the Ramond-Ramond states
and does not break Poincare invariance. This
chiral projection can therefore be identified as the
heterotic string: the superconformal gauge fixed 
$N_s$$=$$(1,0)$ conformal field theory has 
central charge c = (12,8), and all closed string 
states with {\em mixed} left- and right-moving 
worldsheet parity are absent. We have:
\begin{equation}
{\rm type ~ IA}^{\prime} :  (NS+,NS+) ,  ((R+,
NS+) \oplus (NS+ , R+ ))_s
\quad  ,
\label{eq:specihj}
\end{equation}
with bosonic massless particle spectrum as follows:
\begin{equation}
 ({\bf 8_v + 8 })
\times 
{\bf 8_v}  = ({\bf 1,1}) + ({\bf 28 , 1})
+ ({\bf 35 , 1}) + ({\bf 56 , 1} ) + ({\bf 8' , 1})
\quad .
\label{eq:specihsm}
\end{equation}

\vskip 0.1in \noindent
It is helpful to restate our result for the two distinct
freely acting asymmetric orbifolds of the type IIA
string as follows. Modding by a spacetime reflection on a
target space coordinate: $X^9$$\to$$-X^9$,
$\psi^9$$\to$$\psi^9$, 
${\tilde{\psi}}^9$$\to$$-{\tilde{\psi}}^9$, maps the
IIA to the equivalent IIA$^{\prime}$ sum over spin structures. Modding in addition by the
discrete groups, setting $(-1)^{F_L} $$=$$+1$,
$(-1)^{F_R}$$=$$+1$, in {\em both} the Ramond 
and Neveu-Schwarz sectors, defines the truncation to the
type IA$^{\prime}$ orbifold, an anomalous 10d $N_S$=1 theory we will
show can be extended to either of the two ultraviolet
finite and infrared unambiguous, exact renormalized
heterotic string theories. 

\vskip 0.1in \noindent
In closing, it should be noted that our derivation of
the heterotic strings as $N_S$=1 chiral projections of
the 10D type IIA superstring has the following important
consequence: under the chiral projection, the 
zero momentum states in the Hilbert space of the
heterotic descendant, for both physical and ghost 
degrees of freedom, will be unchanged from those
deduced from a BRST analysis of the type IIA 
superstring. In other words, as reviewed in Appendix A,
the measure in the string path integral is unchanged from 
that for the type IIA superstring, namely, preserving the
Wess-Zumino gauge fixed N=(1,1) local worldsheet 
supersymmetry, except for choices of spin structure 
which imply the presence of supermoduli, or conformal 
Killing spinors. Neither is present at genus one, 
except in the Ramond-Ramond sector.

\vskip 0.1in \noindent
Note that the type IIA 
Ramond-Ramond (R-R) p-form potentials are projected
out of the Hilbert space of the descendant, since
the chiral projection 
removes all states in the type IIA Hilbert space 
with mixed left and right worldsheet parity. The
parity projection, however, retains the {\em constant} modes of the odd rank R-R potentials of the type IIA theory.
This suggests that heterotic strings can be
formulated in backgrounds with constant R-R type
IIA pform potentials of definite parity. Note that R-R fluxes, and, consequently,
Dbrane sources, are always absent in the heterotic descendants.

\vskip 0.1in \noindent
Notice that because of our identification of spacetime
parity with the {\em product} of worldsheet parities,
following the projection to positive spacetime parity,
Ramond worldsheet fermions only appear in the 
Hilbert space of the right-moving superconformal 
field theory. We no longer have the ingredients to 
build the spinorial $\bf 8$ or $\bf 8'$ in the Hilbert
space of the left-moving conformal field theory: \footnote{It is conventional to refer to the
superconformal half of the heterotic string theory as right-moving, or anti-holomorphic,
listing the boundary conditions on {\em right-moving} worldsheet fermions 
before those on fermions in the left-moving, or holomorphic, sector \cite{het}: 
([right],[left]), as in the equation above. Thus, it is the right-movers that will provide a
realization of the SO(8) subgroup of the 10D Lorentz group in the heterotic
string theory, giving rise to the $SO(8)_{\rm spin}$ representations listed. Note 
that right-moving worldsheet fields are distinguished by tildes.}

\subsection{Heterotic String Descendants of the Type IIA String}

\vskip 0.1in \noindent An alternative means towards fulfilling the infrared consistency conditions 
on a closed string theory with massless chiral fermions
in the $\bf 8'$ is available for the chiral projection of the type IIA superstring. Notice that
the projection to states with positive spacetime parity eliminates the R-R sector of the
worldsheet superconformal field theory {\em in entirety}. One consequence, of course,
is that the heterotic string theory therefore cannot accommodate Dbranes and,
based on our discussion in the previous section, it is clear that there is no consistent 
extension incorporating open string sectors. 

\vskip 0.1in \noindent
In addition, as mentioned earlier,  the necessary ingredients for building a spinorial 
vacuum no longer exist in the left-moving conformal field theory. We wish to extend
the spectrum of the closed 
string theory in such a way that the low energy limit yields additional supermultiplets
with chiral fermions. Such chiral fermions can contribute 
the necessary compensating terms to the anomaly 
polynomials. The analysis of the hexagon anomaly in the chiral ten-dimensional 
N=1 supergravity reveals that coupling to the chiral fermions of a 10D super-Yang-Mills 
theory with precisely 496 massless vector bosons meets the conditions for the 
cancellation of all anomalies: gauge, gravitational, and mixed. What extension 
to the chiral projection of the type IIA string theory can account for this massless
field content in the low energy field theory limit? Recall that as a consequence of the
chiral projection, we begin with the bosonic massless particle spectrum:
\begin{equation}
(NS+,NS+) \oplus (R+, NS+): \quad ({\bf 8_v + 8 }) \times {\bf 8_v}  = ({\bf 1,1}) + ({\bf 28 , 1})
+ ({\bf 35 , 1}) + ({\bf 56 , 1} ) + ({\bf 8' , 1}) \quad ,
\label{eq:specihsk}
\end{equation}
We wish to augment this bosonic massless particle spectrum with a (${\bf 8_v}$,${\bf 496}$)
of vector bosons, and their ($\bf 8$,${\bf 496}$) chiral superpartners under the
10D N=1 supersymmetry. The two Yang-Mills gauge groups with an adjoint representation
of dimension 496 are $SO(32)$ and $E_8$$\times$$E_8$.

\vskip 0.1in \noindent The clue towards uncovering the nature of the fully consistent heterotic string
lies in the peculiar mismatch in the properties of the Hilbert spaces of 
left and right moving conformal field theories of the IIA string following the chiral 
projection to physical states with positive spacetime parity. Note that the worldsheet 
local superconformal
algebra following the chiral projection remains the familiar (1,1) SCFT underlying the type IIA superstring, except that the superconformal generators of the left-moving (super)conformal field 
theory, which belonged in the ({NS$+$},{R$-$}) sector of the IIA string, can no longer contribute 
to the physical Hilbert space of the heterotic string theories because 
of the restriction to states of positive
spacetime parity!  We emphasize that the total central charge of the worldsheet superconformal 
field theory is (15,15), just as in the type II superstrings, and the transverse degrees of freedom 
remnant after superconformal gauge fixing, following the elimination of timelike and longitudinal
worldsheet (1,1) supermultiplets together with compensating bosonic and fermionic ghosts,
are also the familiar (12,12) of the type II superstrings.\footnote{To the best of our
knowledge, the perspective on the
heterotic string theories offered here is completely new. The traditional approach, including that followed in the 
original papers \cite{het}, has been to invoke the Hamiltonian quantization of independent left-moving and
right-moving 2d conformal field theories with total central charge (15,26): namely, the left-moving half of a 26D bosonic string theory and the right-moving half of a 10D type II superstring. Eight 
of the 24 transverse bosonic LM modes are subsequently paired with the eight transverse RM bosonic modes of the superstring, and identified as the
transverse \lq\lq coordinates" of a 10D target spacetime. Our goal here is to point out that there
exists an alternative derivation of the heterotic string theories as chiral projections of the type IIA
superstring that can reproduce the results of the traditional 
construction. To be precise, 
the {\em physical} Hilbert spaces and {\em on-shell} scattering amplitudes derived in either approach
will be indistinguishable.} Notice that the physical Hilbert space 
has only states of positive definite norm. Thus, without having any impact on the target spacetime
Lorentz and supersymmetry algebra, we can self-consistently extend the {\em left-moving} conformal
field theory with a unitary compact chiral conformal field theory, subject to the overall constraints of 
modular invariance on the expression for the string vacuum amplitude. Recall that invariance of the 
one-loop vacuum amplitude under the modular group of the torus also determines the physical state spectrum.

\vskip 0.1in \noindent From our earlier discussion for the 10D type II superstrings, the one-loop 
vacuum amplitude for a heterotic string theory will therefore take the general form:
\begin{eqnarray}
W_{\rm het} =&& L^{10} (4\pi^2 \alpha^{\prime})^{-5} \int_{\cal F}
\left \{ {{d^2 \tau}\over{4\tau_2^2}} \cdot 
  (\tau_2)^{-4} [\eta(\tau) {\bar{\eta}} ({\bar{\tau}} )]^{-8} \right \}
  \cr
  &&\quad\quad \times 
  {{1}\over{4}}  \left [  ({{{\tilde{\Theta}}_{00}}\over{{\tilde{\eta}}}})^4 - 
  ({{{\tilde{ \Theta}}_{01}}\over{{\tilde{\eta}}}})^4 - ({{ {\tilde{\Theta}}_{10}}\over{{\tilde{\eta}}}})^4 -
  ({{ {\tilde{\Theta}}_{11}}\over{{\tilde{\eta}}}}) ^4 \right ]  \times {\rm Z}_{\rm chiral} (\tau) \quad .
\label{eq:IIh}
\end{eqnarray}
Recall that the factor within curly brackets is already modular invariant. The holomorphic 
sum over Jacobi theta functions is the remnant contribution from the eight transverse
worldsheet fermions of the type IIA string following the projection to states of positive 
spacetime parity. Under a $\tau \to \tau +1 $ transformation, this function transforms as
follows:
\begin{equation}
\tau \to \tau+ 1: ~  \left [ ({{\Theta_{00}}\over{\eta}})^4 - ({{ \Theta_{01}}\over{\eta}})^4 - ({{\Theta_{10}}\over{\eta}})^4 -
  ({{\Theta_{11}}\over{\eta}}) ^4  \right ] \to e^{ 2\pi i /3} \left [ ({{\Theta_{00}}\over{\eta}})^4 - ({{ \Theta_{01}}\over{\eta}})^4 - ({{\Theta_{10}}\over{\eta}})^4 -
  ({{\Theta_{11}}\over{\eta}}) ^4 \right ]  ,
\label{eq:modsh}
\end{equation}
and it is invariant under the transformation $\tau \to -1/\tau$. Thus, the function ${\rm Z}_{\rm chiral}(\tau)$ must transform with the compensating phase
under the $\tau$$\to$$\tau+1$ transformation, and is required to be {\em invariant} 
under a $\tau \to -1/\tau$ transformation. The latter property is very significant. In terms of the
vertex operator construction for the physical states in the chiral conformal field theory, namely,
those counted by the level expansion of the function ${\rm Z}_{\rm chiral}$, it implies 
the self-consistency, or {\em closure}, of the vertex operator algebra. 

\vskip 0.1in \noindent
Closed vertex operator algebras with central charge $c$ $>$ $1$ 
are known to exist for only special values of the central charge. Vertex operator algebras
 with $c$ taking
{\em integer} values are characterized by
the properties of Euclidean, even self-dual lattices of dimensionality $c$ \cite{het}. 
The smallest integer solutions with  $c$ $>$ $1$ are $8$ and $16$, and the 
corresponding euclidean even
self-dual lattices contain, respectively, the direct sum of the root and weight lattices 
of the simply-laced Lie algebras 
$E_8$, and either $E_8$$\times$$E_8$ or $SO(32)$.\footnote{More precisely, the 
even self-dual lattice obtained in the latter case pertains to the Lie algebra
$\rm Spin(32)/Z_2$; the ${\rm Z}_2$ projection removes the
root vectors of squared length unity, so that the massless vector bosons in the string
spectrum live in an ${\bf  496}$ of the simply-laced algebra SO(32).}
Recall that the contribution to
the vacuum energy from a chiral vertex operator algebra with central charge $c$ is $c/24$. 
For $c$$=$$16$, we have $E_{\rm vac}$$=$$ +2/3$, and either of the 
given 16-dimensional euclidean 
lattices contains precisely $480$ lattice-vectors of squared length two. Thus, either
choice is 
a consistent candidate that can provide the $496$ massless vector bosons in the 
(NS+,NS+; ${\bf r}$) = (${\bf 8_v}$, ${\bf 1}$; ${\bf 496}$)
sector, as was required by the infrared consistency conditions mandating the absence of
gauge, gravitational, and mixed anomalies:
\begin{eqnarray}
{{1}\over{4}} \alpha^{\prime} ({\rm mass})_L^2 = N_b^{\mu} 
 + N_b^{I} + N_f^{\mu}  -  1 + \half k_L^2   = {{1}\over{4}} \alpha^{\prime} ({\rm mass})_R^2 =
 {\tilde{N}}_b^{\mu} +  
 {\tilde{N}}_f^{\mu}  -  \half  
\quad .  \label{eq:masshl}
\end{eqnarray}
Here, $\mu$ runs from 1 to 8, labelling the transverse modes of the 10D type IIA
superstring. The index $I$ runs from 1 to 16, labelling the 16 orthogonal directions 
of the euclidean even self-dual lattice that self-consistently extends the left-moving conformal 
field theory, following the chiral projection to type IIA states with positive spacetime parity.
Note that the lattice vector ${\bf k}_L$ has 16 components, and states in the CFT with
${\bf k}_L^2$$=$$2$ correspond to massless physical states in the closed string spectrum.
The function ${\rm Z}_{\rm chiral}(\tau)$ takes the form:
\begin{equation}
{\rm Z}_{\rm chiral} (\tau) = \left [ \eta (\tau ) \right ]^{-16}
       \sum_{ {\bf k}_L \in \Lambda_{\rm 16} }  q^{ \half {\bf k}_L^2} 
\quad , 
\label{eq:latt}
\end{equation}
where $\Lambda_{\rm 16}$ is an even self-dual lattice of rank 16. Namely, for
every pair of vectors ${\bf k}$, ${\bf k}'$ $in$ $\Lambda_{\rm 16}$, $\bf k \cdot k'$ is an even integer,
and both the vector, ${\bf k}$, and its dual, $\bf k^*$, where ${\bf k} \cdot {\bf k^*}$$=$$1$,
belong in the lattice $\Lambda_{\rm 16}$.The transformation $\tau$$\to$$-1/\tau$
simply interchanges the root-lattice with its dual lattice, the direct sum of the weight-lattices
of the irreducible representations of the Lie algebra. The overall multiplicative factor $(-i \tau)^{8}$
is cancelled by the corresponding transformation of the eta function. Under the $\tau$$\to$$\tau+1$ transformation,
the lattice summation instead transforms by an overall phase which is unity for rank 16. Thus,
the only factor of relevance is the overall phase $e^{-2\pi i /3}$ in the transformation of
$[\eta(\tau)]^{-16}$; this phase is cancelled by the corresponding transformation of the
anti-holomorphic sum over spin structures, namely, that for
fermions in the right-moving superconformal
field theory. Invoking boson-fermion equivalence in two dimensions, it is sometimes
convenient to 
write the result for the one-loop vacuum amplitudes of the two heterotic string theories,
respectively, in the alternative form \cite{het}:
\begin{eqnarray}
{\rm Z}_{\rm SO(32)} (\tau) 
=&& {{1}\over{2}} \left [  \left ({ { \Theta_{00} }\over{ \eta }} \right )^{16}
 + \left ({ { \Theta_{01} }\over{\eta }} \right )^{16} + 
 \left ({ { \Theta_{10} }\over{\eta }} \right )^{16} + \left ({ { \Theta_{11} }\over{ \eta }} \right )^{16}
 \right ] \cr 
{\rm Z}_{\rm E_8 \times E_8 } (\tau) 
=&& {{1}\over{4}} \left [  \left ({ { \Theta_{00} }\over{ \eta }} \right )^{8}
 + \left ({ { \Theta_{01} }\over{\eta }} \right )^{8} + 
\left ({ { \Theta_{10} }\over{ \eta }} \right )^{8} + \left ({ { \Theta_{11} }\over{ \eta }} \right )^{8}
 \right ]^2
 \quad , 
\label{eq:latth}
\end{eqnarray}
inferred from the equivalent fermionic representation of the respective chiral   
vertex operator algebras with $c$$=$$16$ by 16 complex (Weyl) worldsheet fermions. 
It is helpful to summarize the full 
massless spectrum of the two heterotic string theories.
For the SO(32) and $E_8$$\times$$E_8$ theories, respectively, we have:
\begin{eqnarray}
&&({\bf 8_v} + {\bf 8}) \times ({\bf 8_v} , {\bf 1}) = ({\bf 1},1) + ({\bf 28},1)+ ({\bf 35}, 1) +
     ({\bf 56} ,1) + ({\bf 8' },1), \quad  {\rm (10D ~N=1 ~ supergravity).}   \cr
 &&({\bf 8_v} + {\bf 8} ) \times ({\bf 1}, {\bf 496})
 =  ({\bf 8_v} , {\bf 496} ) + ({\bf 8 },  {\bf 496}) ,  ~{\rm or},  \cr
 && \quad \quad ({\bf 8_v} + {\bf 8} ) \times ({\bf 1}, {\bf 496})
 =  ({\bf 8_v} , {\bf 120 }, 1 ) + ({\bf 8_v} , 1,{\bf 120 }) + ({\bf 8_v} , {\bf 128 }, 1) +  ({\bf 8_v} , 1, {\bf 128 })
  + ({\bf 8 },  \cdots ) . \cr && 
 \label{eq:pizza}
\end{eqnarray}
We have spelled out the $SO(16)$$\times$$SO(16)$ decomposition of the 
496 states in the adjoint representation of $E_8$$\times$$E_8$ in the last equation.
Note that the $\bf 56$$+$$\bf 8'$ of the 10D N=1 supergravity multiplet is generated by the
${\bf 8}$$\times$${\bf 8_v}$; the gauginos live in an $\bf 8$, precisely as in the type IB
unoriented string, and as required by the anomaly cancellation conditions. However,
unlike the type IB supergravity, there appears to be no consistent extension to the spectrum of 
antisymmetric supergravity potentials in the heterotic string theories because the 
Ramond-Ramond sector of the 
\lq\lq parent" type IIA superstring, namely (R+,R-), has {\em negative} 
spacetime parity.

\vskip 0.3in \noindent
\subsection{Compact Lie Algebras and the R-R Tenform}

\vskip 0.1in \noindent We begin with a simple explanation of the realization of $E_8$$\times$$E_8$$\times$${\rm Z}_2$ in the type IIA string with parallel stacks of 8 D8 branes--- and their 8 orientifold image-branes--- separated by the interval, $X_9^{\prime}$, obtained by T$_9$-dualizing the circle $X_9$ in the type IB superstring with space-filling 16 D9branes, plus their 16 orientifold image-branes, and gauge group O(32). The Dirichlet coordinate separating the O8planes is of length $R_9$. The gauge group realized on each of the D8brane stacks is clearly $O(16)$. Are there any additional massless open string states? If we introduce a D0brane on the stack of D8branes, including its image-D0brane, the D0brane pair can freely explore all of the 16 D8branes plus images. Zero length D0-D0, or D0-D8 strings in the Neveu-Schwarz sector are massive, due to the vacuum energy. However, in the Ramond sector of the open string spectrum, we can describe the vacuum state of the D0brane pair as follows: its intersection with each D8brane is a two-state Ramond vacuum, of charge $\pm \half$, and vacuum energy, $E_0 $$=$$ \half (\half)^2$,

\vskip 0.1in \noindent
In nine dimensions, and below, it is well-known that the two, 
apparently inequivalent, ten-dimensional electrically charged
heterotic string theories with $E_8$$\times$$E_8$ and 
${\rm Spin}(32)/{\rm Z}_2$ gauge symmetry \cite{het}, are, in
fact, related by a T$_9$-duality transformation upon
compactification on a circle.
A Wilson line background in the $E_8$$\times$$E_8$ string
continuously interpolates between the two stable and 
supersymmetric heterotic string vacua, leaving unbroken
all sixteen conserved supercharges. On the other hand,
it was well-known that the only allowed perturbative type IB 
gauge groups
obtained by an analysis of Chan-Paton factors are the 
classical groups; $A_n$, $B_n$, $C_n$, and $D_n$, 
as was proven by Marcus and Sagnotti \cite{marsag}. 
This leaves us with the following puzzle:
by the Polchinski-Witten Type IB-heterotic string-string
duality map, it would have to be true that
the strong coupling dual of the 9D heterotic
$E_8$$\times$$E_8$$\times$$U(1)$ 
vacuum with sixteen unbroken supersymmetries should
be a stable, massless tadpole and tachyon free, {\em nonperturbative}
background of the open and closed unoriented type IB string
\cite{polwit}. It was suggested in the early work \cite{dnotes}, that
in the presence of D0branes, in addition to the 16 D8branes
on either of the two orientifold planes bounding the 
interval in the type IA string with $SO(16)$$\times$$SO(16)$
gauge fields \cite{polwit}, the nonabelian gauge symmetry
might extend to the elusive $E_8$$\times$$E_8$. Many
authors subsequently attempted this problem with partial
success \cite{bachas}, but without pointing out that
the 9d type I $E_8$$\times$$E_8$ vacuum is tachyon and
tadpole free, an exact renormalized background with 16 
conserved supercharges.
I will fill in the gaps in that sketchy presentation in what
follows, in response to questions since put to me, also completing 
the details for type I realizations of all of the
simply-laced, and non-simply-laced, compact Lie algebras
in the Cartan-Weyl classification.

\vskip 0.1in \noindent
My original goal was to provide a realization of the 
exceptional Lie algebras in 9D Dbrane backgrounds of the type IB and type IA strings. 
We will now show that, in fact, Dbranes cover all of the Cartan-Weyl
classification: $A_n$, $B_n$, $C_n$, $D_n$, 
$E_6$, $E_7$, and $E_8$, including both simply-laced, exceptional, and
non-simply-laced Lie algebras. This fact, long elusive, 
becames rather obvious, once we establish a detailed
isomorphism, namely, a one-to-one mapping, 
between the standard root and weight systems of the 
Lie algebras and the sequence of jumps in the ten-form 
when a D0brane crosses a D8brane in a generic type IA 
orientifolds. In addition to reviewing this 
analysis of positive, and negative, vacuum energy contributions from the intersection, or crossing, of 
D0-D8branes, we will give an even simpler derivation, directly in terms of the allowed \lq\lq no-force" configurations of 
D0branes in the worldvolume of the stack of D8branes, such that the state conserves 16 supersymmetries.

\vskip 0.1in \noindent
Recall that, unlike the origin of gauge
symmetry in affine Lie algebras embedded in the bulk
worldsheet conformal field theory, nonabelian gauge symmetry in the 
type IB string open and closed string has a completely
different origin \cite{schwarz,polchinskibook}. The
Chan-Paton wave-functions labeling the endpoints 
of open strings provide a representation of a Lie 
group, rather than Lie algebra, and the massless 
lowest-lying mode in the open string spectrum lives
in the adjoint representation of this group.  This
counting gives rise to what were known to be the list
of possible perturbative type I gauge groups:
$U(n)$, $O(2n)$, $Sp(2n)$, and $O(2n+1)$
\cite{marsag}. 

\vskip 0.1in \noindent We remind the reader that the 
root  and weight lattice and Dynkin diagram representations 
of Lie algebras do not distinguish between the classical and
exceptional algebras in the Cartan-Weyl classification of
the compact Lie algebras. In \cite{flux}, we noticed that a
precise counting of states in the $SO(16)$ spinor weight
lattice is given rather easily by an isomorphism to the
sequence of jumps in the Ramond-Ramond ten-form field
upon D0-D8brane crossings along the interval between
the two O8planes \cite{polwit}. We will derive 
the massless Ramond-Ramond tadpole cancelation
conditions that singles out the stable 9D type IA background 
with $E_8$$\times$$E_8$ nonabelian
gauge symmetry. Note that this is a nonperturbative
background of the type IA, or type IB, string \cite{polwit,dnotes}.

\vskip 0.1in \noindent
Let us begin with type IA backgrounds with 32 D8branes.
The gauge group on the 9D worldvolume of 8 coincident
D8branes, and images, is given by the counting of zero
length strings stretched between any pair of
D8branes on either O8plane. 
Note that for realizations of the classical groups,
we can count massless gauge bosons by the
traditional counting for the (classical) SO(2n) group:
2n choices of D8brane (or image) at one end-point, 
2n-1 choices for the other end-point, factor of two
for symmetry under interchange. This method of
counting is predicted by T$_9$-duality, from the usual
counting of Chan-Paton wavefunctions in the Type
IB vacuum \cite{marsag,schwarz,polchinskibook}.

\vskip 0.1in \noindent
Fortunately, in the $T_9$ dual Type IA backgrounds, we
have an equivalent prescription that extends to cover
exceptional algebras as was discovered by us in 
\cite{flux}: the isomorphism of ten-form profiles
on D8-D8brane crossings to a Lie algebra root lattice.
We include all profile vectors with net vanishing
ten-form background on the O8plane; the compensating
ten-form profile with an overall negative sign exists
for the image D8-D8 crossings. In addition, we impose
both cancellation of dilaton tadpoles and anomalies 
on each O8plane. 
 It is easy to verify the correctness 
of this prescription for the SO(16) lattice when 
$n={\bar{n}}=8$:
\begin{equation}
2n(n-1)/2 + 2{\bar{n}}({\bar{n}} -1)/2 = 56 + 56 \quad
..
\label{eq:rootl}
\end{equation}
We have either (+,+) or (+,-) ten-form profiles for
zero length strings any pair of coincident D8branes.
The factor of 2 counts the negative of the profile, a
factor of 1/2 corrects for overcounting, since this is
an interchange of branes and images: the pair of image
D8branes {\em always} has the compensating ten-form 
background in order that there is no net ten-form
on the O8plane. This is consistent with our requiring
the absence of dilaton tadpoles in the absence of a
gradient in the ten-form at either O8plane. The eight 
gauge bosons transforming in the $U(1)^8$ subalgebra 
arise from the (null) ten-form background due to a
soliton between D8brane to its own image, giving a
total of $120$ massless gauge bosons.

\vskip 0.1in \noindent
Let us move on to the 2mD0-2nD8-O8 background; 
$m$ is an integer, and we will hope for a solution for
n=8, since we wish to keep the $120$ SO(16) gauge 
bosons. The massless R-R tadpole cancellation
for 2mD0-2nD8 strings can be expressed by 
combining the conditions for 2nD8-2nD8, 
2mD0-2nD8, and 2mD0-2mD0 strings stretched
between O8planes. It is obvious that
there is no solution unless m=8, since the O8planes
contribute -16 each:
\begin{eqnarray}
&&(2n)(2n) - 2^4 [ 2n + 2n] +2^8 = 0 \cr
&&(2m)(2n) - 2^4 [ 2m + 2n] +2^8 =0 \cr
&&(2m)(2m) - 2^4 [ 2m + 2m] +2^8 = 0 \quad .
\label{eq:tadpoles1}
\end{eqnarray}

\vskip 0.2in \subsection{Sign of the R-R Tenform and SU(2)
Anomaly Cancellation}

\vskip 0.1in \noindent
The counting of zero length D0-D8 strings on each
O8plane is as follows \cite{flux}. A D0-D8 crossing 
results in soliton string creation, and there are 8
D0-D8 soliton strings, plus images, at each O8plane.
We can represent the ten-form profile at the 8 D0-D8 
crossings by an 8 component vector, the profiles take 
the form:
\begin{eqnarray}
&&( +\half,
+\half,+\half,+\half,+\half,+\half,+\half,+\half )  
\cr
&&  ( +\half,
+\half,+\half,+\half,+\half,+\half,-\half,-\half ) 
\quad \oplus {\rm
permutations} \cr
&&( +\half,
+\half,+\half,+\half,-\half,-\half,-\half,-\half ) 
\quad    \oplus  {\rm
permutations} 
 \quad  .
 \label{eq:charges}
\end{eqnarray}
This gives a total of $2 \cdot ( 1/1! \oplus (8 \cdot
7)/2! ) \oplus (8\cdot 7 \cdot 6 \cdot 5)/4! 2!$ $=$
$128$ vectors. Each ten-form profile corresponds to a 
degenerate vacuum, the distinct profiles differ by
permutations of the 8 D0-D8 solitons, and are 
indistinguishable in the coincidence limit. The
factor of two accounts for the interchange of branes 
and images; the factorials correct for permuting
indistinguishable solitons. Recall the standard
definition of the spinor lattice of O(16): include all
vectors of square length two, with each 
component normalized to $\pm \half$, 
and an {\em  even} total number of plus signs. 
Namely, we have either a (8,0), (6,2), or (4,4), split
of plus and minus signs. The restriction to an odd
total number of plus signs will give the conjugate 
spinor lattice.

\vskip 0.1in \noindent
In the spirit of tracing all string consistency 
requirements to infrared target space physics
\cite{schwarz,polchinskibook}, we require the absence 
of gauge, gravational, and mixed, anomalies. The 
above-mentioned \lq\lq sign" rule on consistent
ten-form profile vectors at the 8 D0-D8 crossings
has a simple low energy field theory origin in the
SU(2) anomaly first noticed in \cite{witsu2}. 
E8 contains SO(16); and
$SO(16) = (SO(4))^4 = (SU(2) \times SU(2))^4$.
The SU(2)'s come in {\em pairs} in all consistent
backgrounds; a single, unpaired, SU(2), in the low 
energy gauge group indicates an anomalous vacuum, 
and it is well-known that the Kalb-Ramond field of 
string theory entering the Green-Scwarz mechanism 
can only correct for an abelian anomaly 
\cite{polchinskibook}. Thus, an infrared consistency
condition is the cancellation of all SU(2) anomalies, 
leading to the following consequence: the D8branes
can only be moved into the bulk spacetime in pairs, 
each with its image. These rules tell us what gauge
groups can arise in non-anomalous vacua by 
moving D8branes in pairs off the O8planes.

\vskip 0.1in \noindent
Note we can invoke a T$_9$ duality transformation,
from type IA with 8 coincident D0-D8
to 8 coincident type IB D1-D9 solitons. How do we deduce
the number of gauge bosons and gauge group for zero
size coincident Dstring-D9 solitons in the T-dual type
IB state? After all, this is a nonperturbative type IB
background, and we need a method other
than the counting of Chan-Paton wavefunctions.
Fortunately, the reversed T$_9$-duality provides
the answer. Recall there is a
Wilson line, $A_9$$=$$( (\half)^8; 0^8)$, responsible
for breaking the original $O(32)$$\times$$U(1)$ to
$O(16)$$\times$$O(16)$$\times$$U(1)$ in the 9D type IB
string, labelling the 32 D9branes as two identical stacks of
sixteen D9branes, coincident with the O9plane. 
Thus, we can deduce via T-duality, the existence of a 9D 
type IB D1-D9 background with sixteen supercharges 
and Yang-Mills gauge group {\em extended to
$E_8$$\times$$E_8$$\times$$U(1)$}; the necessary 
Dstrings are wrapped around the circle. 

\vskip 0.2in \subsection{Affine Lie Algebras and Type I Duals of the CHL Strings}

\vskip 0.1in  \noindent
A related puzzle also having to do with enhanced gauge
symmetry, and with fundamental consequences for the 
connectivity of the Landscape, arises as follows. The
non-simply-laced algebras 
$Sp(2n)$, $SO(2n+1)$, $F_2$, and $G_4$ are known to
arise at  enhanced symmetry points (ESPs) in the CHL
supersymmetry-preserving abelian $Z_N$ orbifold moduli
spaces, each with sixteen supercharges \cite{chl}, including
the 8D $Sp(20)$ and $E_8$$\times$$SO(5)$ 
ESPs in the moduli space of the ${\rm Z}_2$orbifold,
obtained by moding by the order two outer automorphism
exchanging the identical $E_8$ Euclidean self-dual
lattices, accompanied by a ${\rm Z}_2$ shift in the 2D
momentum lattice, and at the fermionic radius \cite{chl}: 
${\bf p}$$=$$(p_L | p_R)$$=$ $( \half , 0 | \half ,
0)$. The resulting shift in masses of string states leaves
massless gauge bosons in the diagonal sub-group
of $E_8$$\times$$E_8$,
while those in the orthogonal $E_8$ acquire masses of
order the string-scale. In 9D and below, the shift vector can be
chosen to preserve target spacetime supersymmetry \cite{chl}.
Thus we have a new 9D half-BPS state with 16 unbroken 
supercharges and $E_8$ gauge group, realized at
level two. Note that there is
no further enhancement of the gauge symmetry at this
radius; if $R$ is tuned to the self-dual radius, the
Kaluza-Klein $U(1)$ current algebra is enhanced to an
$SU(2)$ \cite{chl}.

\vskip 0.1in \noindent
It may be helpful to point out that it is possible to
find sporadic examples of $c $ $>$ $24$
self-consistent holomorphic conformal field theories 
which meet the pre-requisites for the closure and
completeness of the chiral algebra. 
As was shown by J.\ Lykken and S.\--Wei Chung in \cite{cchl}, using
results by E.\ Verlinde, holomorphic
conformal field theories of twisted Majorana fermions
with self-consistent closed operator algebras occur at
only specific values of the central charge, namely, 8, 12,
14, 16, 18, 20, 24, 32, and beyond, deduced by
imposing the requirements of a self-consistent fusion
algebra on the tensor product of an even number of
twisted $ c= \half$ Majorana (real) fermions
\cite{cchl}. Such an analysis cannot provide an exhaustive
classification, but suffices to establish the consistency
of sporadic self-consistent holomorphic CFTs with $c >
24$, a useful complement to lattice classifications.
${\rm Z}_k$.

\vskip 0.1in \noindent
It should be noted that the 
single $E_8$ current algebra is realized at level two 
\cite{chl}. A fermionic realization of an 8D ESP with 
16 unbroken supercharges and gauge group $Sp(20)$ 
was discovered in \cite{chl}. 
It turns out to belong in the $E_8$ moduli space, as
shown by us in \cite{chl}. The full structure of the
moduli spaces, and the intriguing appearance of a 
systematic sequence of electric-magnetic dual enhanced
gauge symmetry points with non-simply laced groups,
was uncovered by us, using the orbifold
technique. This is important, since unlike the simply
laced cases, where electric and magnetic groups are
the same, the electric and magnetic dual groups differ
in the case of any non-simply-laced Lie group. 
$Sp(2n)$$\leftrightarrow$$SO(2n+1)$. It turns out
that it is indeed true that an ESP with non-simply-laced gauge
symmetry can appear, without the magnetic dual ESP, 
in the moduli spaces in \cite{chl} in spacetime dimensions
$9$ $\ge$ $D$ $\ge$ $5$.  Remarkably, precisely as 
required by self-duality of the 4D N=4 supergravity
coupled to super-Yang-Mills gauge theory, it is only in four
dimensions that the moduli 
space contains {\em both} of the necessary enhanced 
symmetry points, with electric-magnetic dual groups 
interchanged. This last observation is due to
J.\ Polchinski.

\vskip 0.1in \noindent
Not surprisingly, we 
discover the $ Sp(20)$ ESP in its moduli space, but in
the T$_9$-dual regime of large type IB radius.
It is quite easy to identify the type IA dual of
9D $E_8$$\times$$(U(1))^2$ moduli space once we
observe the analogy between D0-D8 crossings and
the vectors in the $E_8$$\times$$E_8$ gauge lattice. 
To begin with, it is helpful to write the ten-form
vectors for 8 D0-D8 crossings, and their 8 image
crossings, as 16 component profile vectors. Label the 
slots in the 16 component profile vectors as follows: 
$(1,2,3,4,5,6,7, 8| {\bar{1}}, {\bar{2}}, {\bar{3}},{\bar{4}}, 
{\bar{5}}, {\bar{6}},{\bar{7}}, {\bar{8}})$. It can be
confirmed that this global pairing of D0-D8branes, and
images, is compatible with all 128 ten-form profiles 
listed above, now written in a 16-component basis.
There is no new information in the last 8 slots
of these vectors; they are the negatives of the first
8:
\begin{eqnarray}
&&( +\half,
+\half,+\half,+\half,+\half,+\half,+\half,+\half | -
\half,
- \half, - \half, - \half, - \half, - \half, - \half,
- \half)  
\cr
&&  ( +\half,
+\half,+\half,+\half,+\half,+\half,-\half,-\half | -
\half,
- \half,- \half, - \half, - \half, - \half, + \half, +
\half) 
 \cr
&&( +\half,
+\half,+\half,+\half,-\half,-\half,-\half,-\half |  -
\half,
- \half,- \half, - \half, + \half, +  \half, + \half,
+ \half) 
 \quad  ,
 \label{eq:dcharges}
\end{eqnarray}
plus all vectors equivalent up to permutations of the
first 8 components. Is it possible to find additional 
sets of ten-form profile vectors that meet infrared 
consistency, namely, dilaton tadpole cancelation
and the absence of SU(2) anomalies, but {\em without} 
a global mutually compatible pairing of all 8 D0-D8 
crossings and 8 image D0-D8 crossings?

\vskip 0.1in \noindent
For readers familiar with the realization of Lie
algebras
by Majorana fermions--- the worldsheet framework
for the fermionic ESPs in the moduli spaces in
\cite{cchl,chl}, it should be obvious that
many solutions to this problem are already known. 
The minimal block of 2n-component vectors which 
does not admit a mutually compatible pairing, or 
complexification, has n=8, as was proven in
\cite{cchl}:
\begin{eqnarray}
&&( +\half,
+\half,+\half,+\half,+\half,+\half,+\half,+\half | -
\half,
- \half, - \half, - \half, - \half, - \half, - \half,
- \half)  
\cr
&&  ( +\half,
+\half,+\half,+\half,-\half,-\half,-\half,-\half | +
\half,
+ \half, + \half, + \half, - \half, - \half, - \half,
- \half) 
 \cr
&&  ( +\half,
+\half,-\half,-\half,+\half,+\half,-\half,-\half | +
\half,
+\half, -\half, - \half, + \half, +\half, - \half, -
\half) 
 \cr
 &&( +\half,
-\half,+\half,-\half,+\half,-\half,+\half,-\half | 
+\half,
- \half,+ \half, -\half, + \half, - \half, + \half, -
\half) 
 \quad  .
 \label{eq:rcharges}
\end{eqnarray}
The analogy between 8 branes, and 8 images, and 
16 Majorana worldsheet fermions is as follows. The
Ramond vacuum of a Majorana fermion exists in one 
of two possible states which we denote $\pm \half$. In
the worldsheet framework underlying the 
fermionic ESPs of the moduli spaces in \cite{chl},
the vacuum amplitude is a sum over sectors with distinct
Ramond, or Neveu-Schwarz, boundary condition
for the 32 worldsheet fermions in the bosonic CFT
with total central charge 16. A mutually compatible
pairing of Majorana fermions among all sectors summed
in the vacuum amplitude provides a complexification of
Majorana fermions, $\psi^i + i \psi^{{\bar{i}}}$, $i=1,\cdots , n$, 
where $n$$\le$$16$. Such a complexification gives $n$
complex fermions, each with central charge one, and,
$n$ U(1)s in the gauge group, since all $n$ lowest
excitations in the NS vacuum are retained in the 
string spectrum: $\psi^i_{-1} \psi^{{\bar{ i}}}_{-1}
|0>$.

\vskip 0.1in \noindent
Conversely, if a complexification of 2n worldsheet 
Majorana fusion algebras does not exist,
the gauge group in the type IA vacuum will have
$n$ fewer U(1)'s. It was proven in \cite{cchl} that the minimal 
solution has $n$ $=$ $8$. Moreover, that the basic 
rules for the overlap of common signs among vectors 
in the block of ten-forms listed above originate as 
follows: any pair of profile vectors is required have
an overlap of 0 mod 4, while any triad must have 
overlap 0 mod 2. Both of these conditions originate in
the ambiguity in the fusion rules of a 2d Majorana fermion
conformal field theory; the full
derivation can be found in \cite{cchl}. 

\vskip 0.1in \noindent
The 9D Type IA string with sixteen unbroken
supercharges but eight fewer U(1)'s is a new stable
half-BPS state; we introduce the ten-form profiles
above in (\ref{eq:dcharges}), and (\ref{eq:rcharges}),
O8planes at the two endpoints of the interval. The
tadpole cancelation conditions are identical to those
in the previous section. The absence of a global
pairing of 8D0-D8 crossings, and 8 image crossings, 
on either O8plane implies a gauge group with 8 fewer
U(1)s. The 9D gauge group is $E_8$$\times$$U(1)$.
This theory is the type IB dual of the 9D heterotic 
CHL string inferred in \cite{chl} as an asymmetric orbifold.
In 8D, it contains both $E_8$$\times$$SO(5)$
and $Sp(20)$ ESPs \cite{chl}. 

\vskip 0.1in \noindent
Our identification of an isomorphism between the
Ramond-Ramond tenform field in the bulk between the
O8planes and the root and weight lattices of a Lie algebra 
in the full Cartan-Weyl classification also leads to a rule for
the {\em sign} of the tenforms, necessitated by
cancellation of all SU(2) anomalies in the generic
D0-D8-O8 background. Finally, we make the following
important observation. The type IB--heterotic duality map with $R_{\rm H}$ set to the
Dirac fermion radius also establishes that the fermionic CHL strings \cite{chl},
and type IA--IB duals \cite{flux}, 
are exact renormalized conformal field theory backgrounds describing
 {\em weakly-coupled} ESPs in both the heterotic and 
the weak-strong dual type IA-IB string moduli spaces, in nine and lower target 
spacetime dimensions \cite{flux}. As pointed out in the main text, while the
type IB dual is strongly coupled, a T-duality gives a
type IA string which is weakly-coupled, so long as the 
heterotic string does not approach the 
self-dual compactification radius, $R_{\rm H} = \alpha^{\prime 1/2}$ \cite{polwit}.  We evade
this regime by matching the normalization of the heterotic and type IA string
vacuum functionals in the small volume, sub-string-scale, weakly-coupled, 
regime of the type IA string, restricting the compactification radii of the dual 
O(32) heterotic string to the large volume regime, 
$R_{\rm H} >> \alpha^{\prime 1/2}$, $R_{\rm IB} >> \alpha^{\prime 1/2}$.

\end{document}